\documentclass{emulateapj}
\usepackage{graphicx}
\usepackage{epsfig}
\usepackage{natbib}
\newcommand{\eg}{{\rm e.g.\ }}

\newcommand{\cm}{{\rm\thinspace cm}}
\newcommand{\km}{{\rm\thinspace km}}

\newcommand{\pcc}{\hbox{$\cm^{-3}\,$}}

\newcommand{\s}{{\rm\thinspace s}}

\newcommand{\erg}{{\rm\thinspace ergs}}
\newcommand{\ps}{\hbox{\s$^{-1}\,$}}

\newcommand{\ergps}{\hbox{$\erg\s^{-1}\,$}}

\newcommand{\kmps}{\hbox{$\km\s^{-1}\,$}}
\newcommand{\pcmsq}{\hbox{$\cm^{-2}\,$} }

\newcommand{\halpha}{H$\alpha$}

\newcommand{\hi}{H{\sc i}}
\newcommand{\hii}{H{\sc ii}}
\newcommand{\nii}{[N{\sc ii}]}
\newcommand{\oiii}{[O{\sc iii}]}

\newcommand{\nH}{\hbox{$N_{\rm H}$}}

\newcommand{\Msol}{\hbox{$\thinspace M_{\sun}$}}
\newcommand{\Zsol}{\hbox{$\thinspace Z_{\sun}$}}

\shorttitle{Superbubble X-ray Energy Budgets}
\shortauthors{Jaskot, Strickland, Oey, Chu \& Garc\'{i}a-Segura}

\begin{document}

\title{ Observational constraints on superbubble X-ray energy budgets}

\author{A. E. Jaskot, \altaffilmark{1}
	D. K. Strickland,\altaffilmark{2}
	M. S. Oey,\altaffilmark{1}
	Y.-H. Chu,\altaffilmark{3} and
	G. Garc\'{i}a-Segura.\altaffilmark{4}
	}	

\altaffiltext{1}{University of Michigan, Dept. of Astronomy, 
        830 Dennison Bldg., Ann Arbor, MI 48109, USA.}
\altaffiltext{2}{Department of Physics and Astronomy,
        The Johns Hopkins University,
        3400 N.~Charles St., Baltimore, MD 21218, USA.}
\altaffiltext{3}{University of Illinois, Dept. of Astronomy, Urbana, 
	IL 61801, USA.}
\altaffiltext{4}{Instituto de Astronom\'{i}a-UNAM, Apartado Postal 877,
        Ensenada, 22800 Baja California, M\'{e}xico.}

\begin{abstract}
The hot, X-ray-emitting gas in superbubbles imparts energy and enriched material to the interstellar medium (ISM) and generates the hot ionized medium, the ISM's high-temperature component. The evolution of superbubble energy budgets is not well understood, however, and the processes responsible for enhanced X-ray emission in superbubbles remain a matter of debate. We present {\it Chandra} ACIS-S observations of two X-ray-bright superbubbles in the Large Magellanic Cloud (LMC), DEM L50 (N186) and DEM L152 (N44), with an emphasis on disentangling the true superbubble X-ray emission from non-related diffuse emission and determining the spatial origin and spectral variation of the X-ray emission. An examination of the superbubble energy budgets shows that on the order of 50\% of the X-ray emission comes from regions associated with supernova remnant (SNR) impacts. We find some evidence of mass-loading due to swept-up clouds and metallicity enrichment, but neither mechanism provides a significant contribution to the X-ray luminosities. We also find that one of the superbubbles, DEM L50, is likely not in collisional ionization equilibrium. We compare our observations to the predictions of the standard Weaver et al. model and to 1-D hydrodynamic simulations including cavity supernova impacts on the shell walls. Our observations show that mass-loading due to thermal evaporation from the shell walls and SNR impacts are the dominant source of enhanced X-ray luminosities in superbubbles. These two processes should affect most superbubbles, and their contribution to the X-ray luminosity must be considered when determining the energy available for transport to the ISM.  

\end{abstract}

\section{Introduction}

The hot phase ($T \ga 10^{6}$ K) of the interstellar medium (ISM)
is generated primarily by the mechanical power imparted by stellar
winds from massive stars and supernovae (SNe). 
The clustered nature of massive star formation necessarily dictates that
some stellar winds and core-collapse
SNe will occur in low-density cavities excavated
by other winds or SNe, thus creating superbubbles. 

Theoretical work by \citet{maclow88} demonstrated that superbubbles could be 
considered more powerful versions of stellar wind-blown bubbles, which
already had a well-developed theoretical model 
\citep{dyson73,castor75,weaver77}. The most detailed of these papers was
that of \citet{weaver77}, which has become the de-facto standard model
explaining the structure and physics of wind-blown bubbles and superbubbles.

Superbubbles play important roles in generating and maintaining the 
multi-phase ISM in star-forming galaxies 
\citep[see \eg][]{norman89,norman96,oey97}.
The influence of superbubbles can extend beyond the thin gaseous disk,
as some superbubbles vent their hot, metal-enriched plasma into the
galactic halo, creating galactic fountains \citep{bregman80}
in normal spiral galaxies like the Milky Way and feeding galactic superwinds
in starburst galaxies \citep{ham90}.

As the hot, collisionally ionized plasma in superbubbles emits primarily by
X-ray line emission and thermal bremsstrahlung, X-ray observations provide
a crucial window on these energetic phenomena. However both X-ray and
optical observational studies indicate that our current
understanding of superbubble physics and energetics is incomplete.

The earliest imaging X-ray observations of star-forming complexes 
in the Magellanic Clouds with the {\it Einstein Observatory} 
revealed that some superbubbles were indeed
detected as X-ray sources \citep{chu90,wang91}. Surprisingly, the
detected superbubbles typically had X-ray luminosities 
an order of magnitude greater than predicted from the standard Weaver et al. model
(we refer to these as X-ray-bright superbubbles). These X-ray-bright superbubbles demonstrate that the processes responsible for X-ray emission in superbubbles are not yet fully understood.
Not all superbubbles are over-luminous, however.
Other superbubbles in the Large Magellanic Cloud (LMC) are as faint or fainter than expected
based on the Weaver et al. model (\citealt{chu95}; referred to as 
X-ray-dim superbubbles). Subsequent observations using the {\it R\"{o}ntgensatellit} ({\it ROSAT}) and {\it ASCA} X-ray Observatories \citep[see \eg][]{chu93,magnier96,dunne01} have confirmed these results.

Solutions to the X-ray-bright superbubble problem fall into two categories; (1) replacing the
standard Weaver et al. model at some or all epochs of superbubble growth, and 
(2) supplementing and extending the standard Weaver et al. model.

A popular and elegant example of the first approach was pioneered by
\citet{chu90}, who proposed that the impact of the blast waves from
any SNe that occurred near the superbubble shell can shock-heat the
shell to X-ray-emitting temperatures. Using 1-D models employing the
Kompaneets thin shell approximation \citep{kompaneets60} 
they demonstrated that this shock-heating
leads to a large, but \emph{temporary}, increase
in the total X-ray luminosity from the superbubble.
Evidence exists for supernova remnant (SNR) shocks within superbubbles, and SNR shell impacts
are also the most plausible cause of 
the unusually high shell expansion velocities seen in some superbubbles
\citep{chu94,oey96c}. Nevertheless, their precise contribution to the X-ray luminosity is uncertain. 

The most-promising example of a solution of the second kind is that of
\citet{silich01}. The standard Weaver et al. wind-blown bubble model takes no account
of metal enrichment once SNe occur; the metal abundances in the hot
plasma are assumed to be the same value at all times.
For a collisionally ionized plasma in the
temperature range $T = 10^{6}$ --- $10^{7}$ K the X-ray emissivity
is roughly proportional to the metal abundance of the plasma, as line
and recombination dominate over pure thermal bremsstrahlung.
 \citeauthor{silich01} point out that
if the plasma is enriched in abundance by a factor of 10 or more 
by SN activity, then a similar enhancement in the soft X-ray luminosity
should be expected. In this hypothesis the structure of the 
hot X-ray-emitting superbubble interior 
remains that given by the Weaver et al. model  
at all times, but its chemical enrichment must be
taken into account self-consistently. 
This hypothesis remains to be tested -- as yet 
there is no observational evidence for high 
metal enrichment in X-ray-bright superbubbles.

Constraining the extent to which these, or any other, mechanisms contribute to the total X-ray emission of superbubbles requires higher quality X-ray data than was available with the earlier
generations of X-ray observatories 
({\it Einstein}, {\it ROSAT}, and {\it ASCA}). Potential discriminants
between the different models hinge on spatial variations in the X-ray spectral
properties of the plasma within superbubbles, which requires observations with an instrument
combining high spatial resolution, moderate to high spectral resolution and
high sensitivity. Both modern X-ray observatories, the 
{\it Chandra X-ray Observatory} and the {\it XMM-Newton}, fulfill these
requirements.

We have obtained and analyzed {\it Chandra}
observations of two X-ray-bright LMC superbubbles,
with the Advanced CCD Imaging Spectrometer (ACIS) instrument. The target objects are DEM L50 and DEM L152 from the\citet{davies76} \halpha~catalog of LMC \hii~regions.  DEM L152 is
also part of the nebula cataloged as N44 by \citet{henize56}.
Other properties of these superbubbles are highly
constrained with existing multi-wavelength observations, properties such as the host stellar population, shell expansion
characteristics, local interstellar gas density and intrinsic absorption. We adopt a distance 
of 50 kpc to the LMC \citep{freedman01}. At this distance an angular offset of $1\arcsec$
corresponds to a physical size of 0.24 pc.

DEM L50 has not previously been the target of pointed X-ray
observations. Optical and \hi~observations are reported in \citet{oey02}.
It is a roughly elliptical bubble with major and minor axes of 
$9\arcmin$ and $7\arcmin$ respectively, associated with the superimposed
SNR N186 D on its northern edge. It is unclear whether the SNR is 
physically connected to the superbubble or whether this is a
chance projection of separate objects along a line of sight. Otherwise DEM L50
is relatively isolated from other \hii~nebulae and young stellar clusters.

In contrast DEM L152 is part of the large and relatively crowded
N44 complex, a set of shells and \hii~regions
covering $\sim 25\arcmin$ of sky \citep{meaburn91}. 
Given the limited $8\arcmin$ field of view of
individual {\it Chandra} ACIS CCD chips we choose to concentrate on
the X-ray-bright features Shell 1, the South Bar, and Shell 3 that
were identified in earlier {\it ROSAT} and {\it ASCA} observations \citep{chu93,magnier96}.

By exploiting {\it Chandra}'s $\sim 1 \arcsec$ spatial resolution
we separate the true superbubble X-ray emission from non-related 
diffuse X-ray emission in the immediate vicinity of each superbubble and
investigate the degree of spatial variations in soft X-ray properties
within these bubbles (see \S~\ref{sec:spectra}). If the diffuse X-ray
emission in these bubbles is very highly structured, arising in
small-scale regions associated with the optical shells, then these 
{\it Chandra} observation have the spatial resolution to detect this.
In \S~\ref{sec:models}
we compare our results to the predictions of the standard 
\citeauthor{weaver77} model and to 1-D hydrodynamical simulations including
SN-shell impacts and discuss superbubble energy budgets in light of our observations.
We discuss the metallicity-enhancement scenario and the possibility of mass-loading from clouds in light of our observations in \S~\ref{sec:discussion}.

\section{Observations}
\label{sec:reduction}

DEM L152 was observed with the {\it Chandra} ACIS-S instrument on
2002 September 22 for $\sim 20$ ks (observation identification number 
[ObsID] 3356), and DEM
L50 was observed for $\sim 40$ ks on 2003 January 01 
(ObsID 3355, see Table.~\ref{tab:obslog}). The observations were made 
 to accommodate the majority of the optical nebula of each bubble
on the more sensitive back-illuminated ACIS S3 CCD chip, but parts of each
nebula spill onto the adjacent S2 and S4 CCD chips.

{
\begin{deluxetable*}{llllrll}
\tabletypesize{\scriptsize}%
\tablecolumns{7}
\tabletypesize{\small} %
\tablewidth{0pc}
\tablecaption{{\it Chandra} observations of DEM L50 and DEM L152
        \label{tab:obslog}}
\tablehead{
\colhead{Target}    & \colhead{R.A.} & \colhead{Decl.}
  & \colhead{Date}  & \colhead{Instrument} 
  & \colhead{ObsID} & \colhead{Exposure time (CCD chip)}
}
\startdata
DEM L50  & $04^{h}59^{m}47^{s}$ & $-70\degr11\arcmin37\arcsec$ &
2003-01-01 & ACIS-S & 3355  & 37011.9 (S3), 36720.9 (S2), 36714.5 (S4) \\
DEM L152 & $05^{h}22^{m}17^{s}$ & $-67\degr56\arcmin38\arcsec$ &
2002-09-22 & ACIS-S & 3356  & 18713.6 (S3), 18713.6 (S2), 18713.6 (S4) \\
\enddata
\tablecomments{The right ascension and declination values quoted (J2000.0 
  coordinates) are the approximate center of the superbubble cavity as seen
  in \halpha~images. The ObsID is the unique observation identification number
  assigned by the {\it Chandra} Science Center.
  The exposure times (in seconds) quoted are those remaining after
  the removal of periods of higher-than-normal background. The specific 
  ACIS CCD chip to which the exposure time applies is noted in parentheses.
}
\end{deluxetable*}}

Data reduction and analysis were performed using the {\it Chandra} CIAO
software package (version 3.4) with the associated calibration database
(CALDB version 3.3). In addition HEASOFT (version 6.1.1) was used for
some tasks, including spectral fitting with XSPEC (version 12.5.1).

Each raw data set was reprocessed to take account of the latest calibration
of important effects such as 
time-dependent CCD gain, charge transfer inefficiency and contamination
of the optical-blocking filter. Periods during the observation that experienced
higher-than-normal levels of Solar X-ray and particle events (commonly referred
to as background flares) were removed
using the iterative $3\sigma$ clipping method described in 
\citet{strickland04a}. The effective on-source exposure time remaining in
each ACIS CCD chip after this procedure is given in Table~\ref{tab:obslog}.
Only a small fraction, $\la 1$\%, of the total exposure time was lost to
background flares in either observation. This is much lower than
the typical $\sim 20$\% fraction of ACIS observation time affected by flares 
\footnote{See ``General discussion of the quiescent and flare components of the ACIS background", by Maxim Markevitch (2001) at \url{http://cxc.harvard.edu/contrib/maxim/bg/index.html}.}.

We used the wavelet-based source detection algorithm {\sc wavdetect}
to search for point-like objects in each X-ray observation. Each chip
was treated separately, searching for sources in images created in
the soft X-ray 0.3 -- 2.0 keV energy-band, hard X-ray 2.0 -- 8.0 keV
energy band and total 0.3 -- 8.0 keV energy band. Only sources with
signal-to-noise ratios $\ge 2$ were accepted as point sources
in the initial iteration of the data analysis.
In a few cases,
low S/N features that most probably are genuine point sources
were missed by the source-detection algorithm. These sources became apparent
only after all brighter point sources had been removed. We assessed
each such feature, and removed those that appeared to be point-like based on
our personal scientific judgment. The point sources are shown marked on 
soft ($E=0.3$ -- 2 keV) and hard 
($E=2$ -- 8 keV) energy band images of DEM L50 and DEM L152
in Fig.\ref{fig:point_sources}.

{
\begin{figure*}[!ht] 
\epsscale{1.0}
\plotone{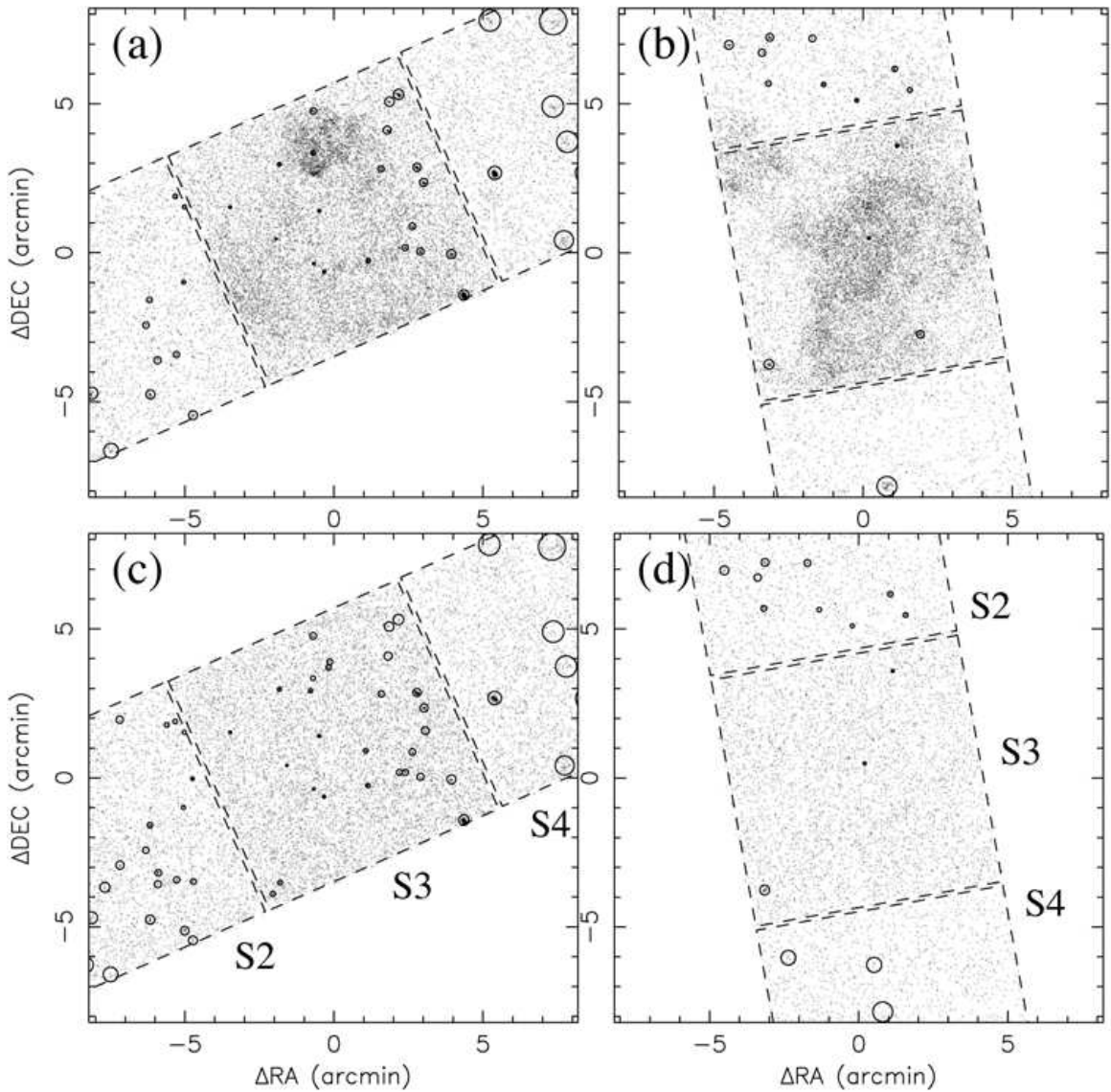}
  \caption{X-ray emission in the vicinity of DEM L50 (a and c) and DEM L152
  (b and d). Each image
  shows the raw, unsmoothed, photon distribution in the soft
  0.3--2.0 keV (panels a and b) and and hard 2.0--8.0 keV
  energy bands (panels c and d), binned in $0\farcs984$-wide
  pixels, over a $\sim 240 \times 240$ pc ($16\farcm4 \times16\farcm4$)
  region centered on the coordinates given in Table~\ref{tab:obslog}.
  Point-like X-ray sources detected in the soft and hard
  energy bands are shown surrounded by a circle, equivalent in size
  to the region used to remove the point sources from the images
  and spectra. The images are displayed using a asinh intensity scale.
  }
  \label{fig:point_sources}
\end{figure*}}

We screened out the events from detected point sources in the data 
used for both imaging and spectral analysis of the superbubbles 
to a radius equivalent to 4 Gaussian $\sigma$, based on a fit to the
radius of the {\it Chandra} PSF as a function of off-axis angle.
The holes left in any image by source removal were filled in using the
CIAO task {\sc dmfilth}. 
The observed distribution of pixel
values in a background annulus of thickness 6 ACIS pixels ($\sim 3\arcsec$)
around the point source defines
the probability  density function from which random values
were chosen to fill in the 
source region (The {\sc poisson} option in {\sc dmfilth}). 
Care was taken to ensure that the background annulus chosen 
did not contain other point sources which would bias the interpolation.

For spectral analysis, sources detected in any of the three bands (the soft,
hard and total bands described above) were excluded when creating
diffuse-emission-only spectra. The soft band source lists were used for any image of the diffuse emission created within the
energy range 0.3 -- 2.8 keV. The hard and total band source lists were used
for any image created within the 2.8 -- 8.0 keV energy band. We discuss the data reduction of the images in \S~\ref{sec:images:data} and the data reduction of the spectra in \S~\ref{sec:spectra:data}.

\begin{figure*}[!ht]
\epsscale{1.0}
\plotone{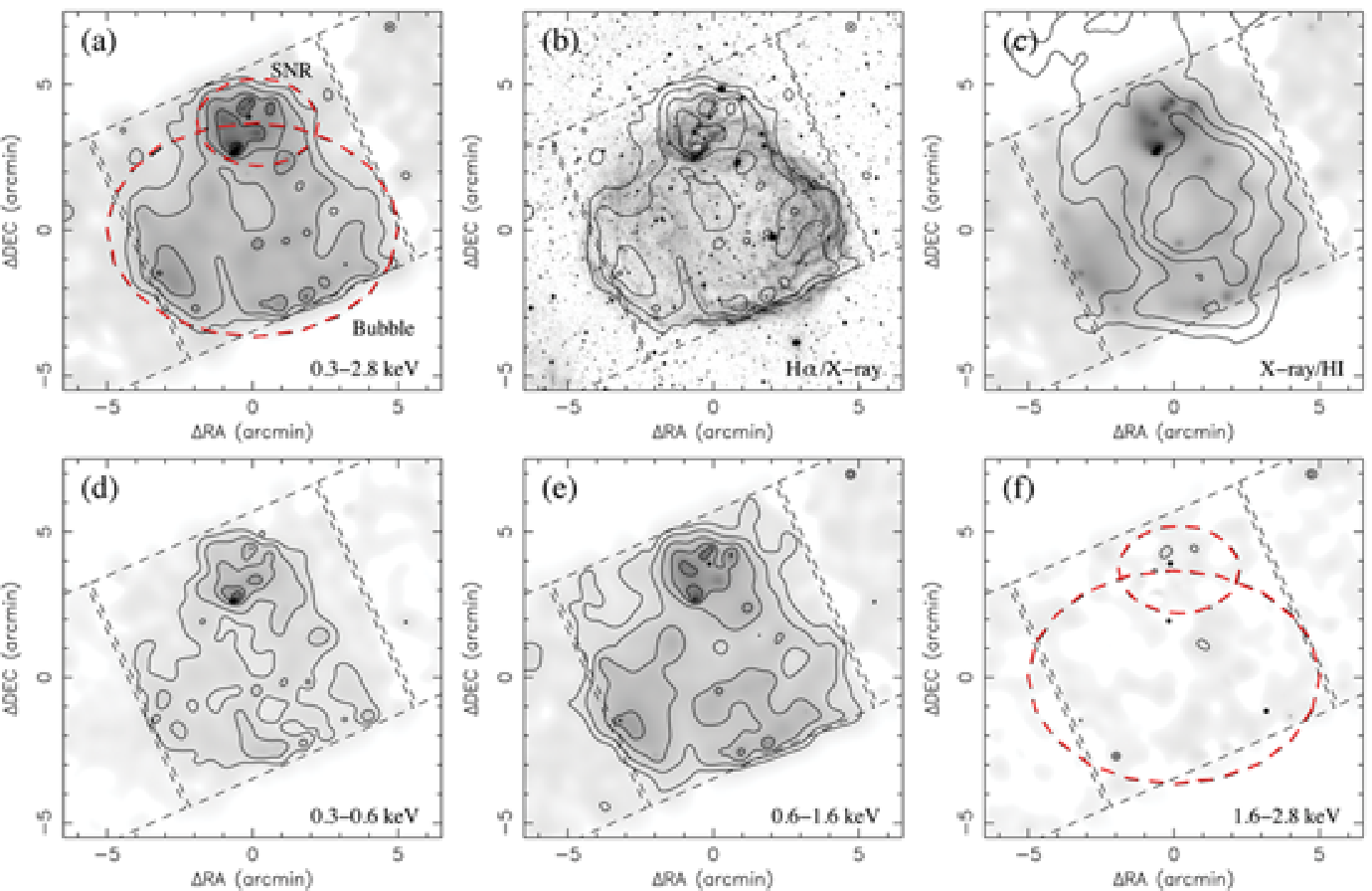}
  \caption{Adaptively-smoothed background-subtracted 
  images of the diffuse X-ray emission
  associated with DEM L50. 
  Panel (a) shows the $E=0.3$ -- 2.8 keV emission on a square root 
  intensity scale, overlaid with contours increasing in factors of two 
  beginning at a surface
  brightness of $\Sigma_{X} = 4 \times 10^{-7}$ counts s$^{-1}$ 
  arcsec$^{-2}$.
  These contours are shown overlaid on a narrow-band \halpha~image from \citet{oey02} in (b). 
  Panel (c) shows the $E=0.3$ -- 2.8 keV emission in grey scale overlaid
  with contours of \hi~column density from \citet{oey02} linearly spaced between
  $\nH = 0.5 \times 10^{21} \pcmsq$ and $2 \times 10^{21} \pcmsq$.
  Panels (d), (e) and (f) show adaptively smoothed X-ray images in the
  $E=0.3$ -- 0.6, $E=0.6$ -- 1.6 and $E=1.6$ -- 2.8 keV energy bands
  respectively, using a square root intensity scale. 
  Contour levels increase in factors of two 
  beginning at a surface
  brightness of $\Sigma_{X} = 2 \times 10^{-7}$ counts s$^{-1}$ 
  arcsec$^{-2}$. Dashed lines denote the edges of the {\it Chandra} ACIS 
  CCD chips, and the dashed red lines in (a) and (f) show the spectral extraction regions.}
  \label{fig:img_deml50}
\end{figure*}


\section{Image Analysis}
\label{sec:images}

\subsection{Image Data Reduction}
\label{sec:images:data}

For the purposes of background subtraction we used 
the blank-sky datasets provided as part of the {\it Chandra} CALDB, as
the diffuse soft X-ray emission from both DEM L50 and
DEM L152 covers a large fraction of the S3 chip in each observation. 
These blank-sky data sets comprise hundreds of kiloseconds of
data per CCD chip, 
providing a high S/N estimate of the typical background experienced
by the ACIS detectors. 
Since the net background experienced in the ACIS detector 
does change with time, the 
blank sky data was re-normalized so that the
mean X-ray surface brightness (excluding point sources) for each
chip in the $E=3$ -- $6$ keV energy band matched that in each
observation. Typical variations in the hard X-ray background in
the ACIS instrument 
are  $\sim 2$\% (root mean square) from observation to observation
\citep{hickox06}. 

\begin{figure*}[!ht]
\epsscale{1.0}
\plotone{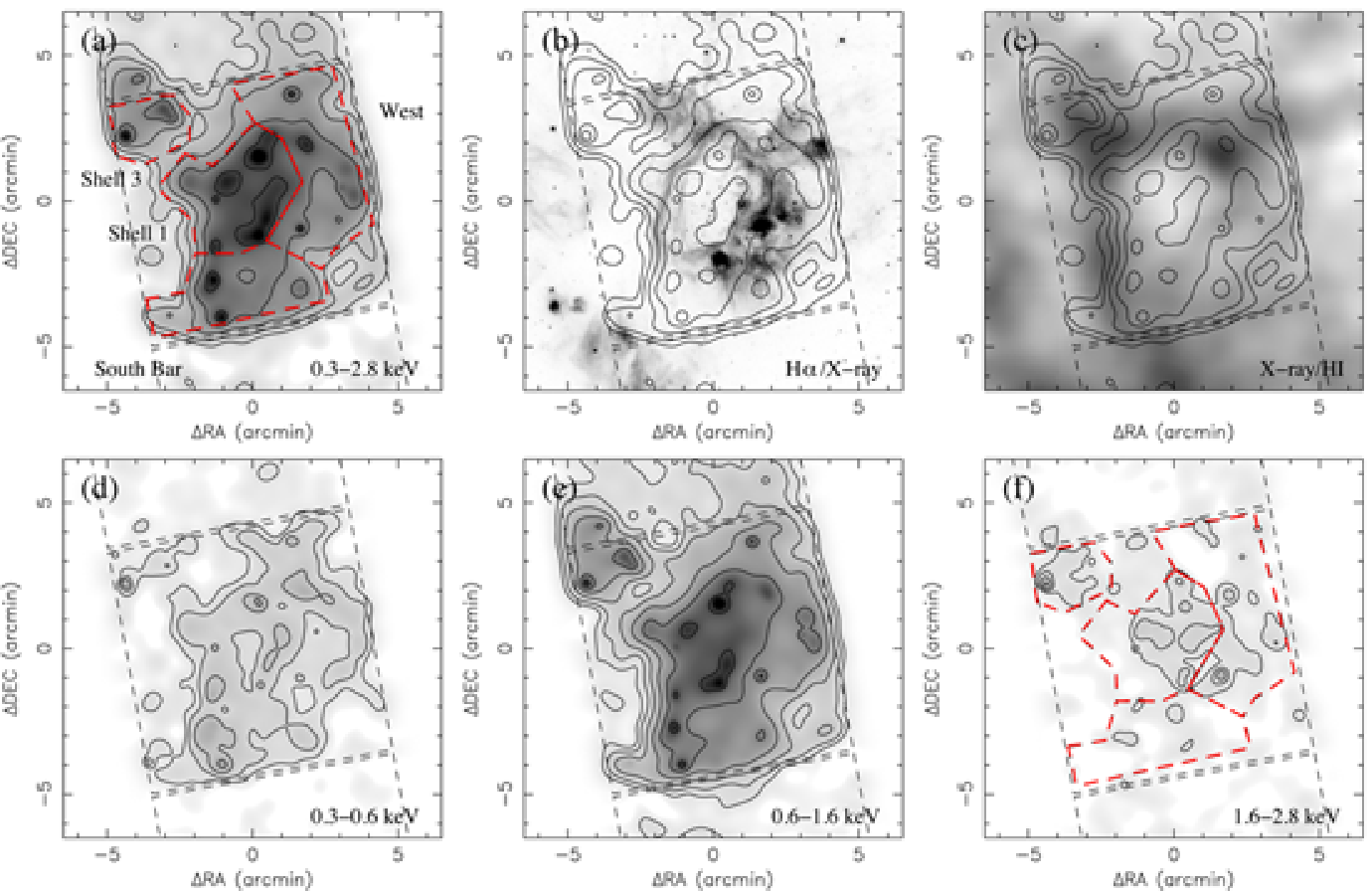}
  \caption{Adaptively-smoothed background-subtracted 
  images of the diffuse X-ray emission
  associated with DEM L152. 
  Panel (a) shows the $E=0.3$ -- 2.8 keV emission on a square root 
  intensity scale, overlaid with contours increasing in factors of two 
  beginning at a surface
  brightness of $\Sigma_{X} = 4 \times 10^{-7}$ counts s$^{-1}$ 
  arcsec$^{-2}$.
  These contours are shown overlaid on a narrow-band \halpha~image from the Magellanic Emission Line Survey (MCELS; \citealt{smith99b}) in (b). 
  Panel (c) shows the \hi~column density from \citet{kim03}, linearly spaced between
  $\nH = 2 \times 10^{21} \pcmsq$ (white) and $6 \times 10^{21} \pcmsq$
  (black), overlaid with the  $E=0.3$ -- 2.8 keV emission contours from (a).
  Panels (d), (e) and (f) show adaptively smoothed X-ray images in the
  $E=0.3$ -- 0.6, $E=0.6$ -- 1.6 and $E=1.6$ -- 2.8 keV energy bands
  respectively, using a square root intensity scale. 
  Contour levels increase in factors of two 
  beginning at a surface
  brightness of $\Sigma_{X} = 2 \times 10^{-7}$ counts s$^{-1}$ 
  arcsec$^{-2}$. 
  Dashed lines denote the edges of the {\it Chandra} ACIS 
  CCD chips, and dashed red lines in (a) and (f) show the spectral extraction regions.}
  \label{fig:img_deml152}
\end{figure*}

We used a hard X-ray band to calculate the renormalization factor for 
the blank-sky data sets as soft diffuse X-ray emission is present
in both the S2 and S3 chips in the two observations 
(see Fig.~\ref{fig:point_sources}). The majority of the
superbubble emission falls within the field of view of the S3 chip
in both observations. Part of DEM L50 spills over onto 
the S2 chip, however, and N44 shell 3 (a SNR; see \citealt{chu93}) is a strong soft
X-ray source in the S2 chip of the DEM L152 observation. Excess
soft X-ray emission is not apparent on the S4 chip in either observation.

Note that apparently diffuse hard
X-ray emission in a superbubble has only been detected in 30 Doradus C \citep{smith04,yamaguchi10}.
Our method of re-normalizing the blank-sky data sets means that
there will be no significant diffuse X-ray emission in the
$E=3$ -- 6 keV energy band, once point sources and the background
have been subtracted. This approach is valid, in that a visual
inspection of the raw, non-background-subtracted data reveals no
evidence for hard X-ray emission within DEM L50 or DEM 152, and
the mean source-free hard X-ray surface brightness
in each chip of our observations is less than or equal to that in the blank-sky data. We cannot rule out the presence
of diffuse hard X-ray emission in our observations of DEM L50 and DEM L152,
but if such emission is present, it is very faint.

For DEM L50, the S3 chip needed essentially no renormalization, as the observed source-free  $E=3 - 6$ keV surface brightness 
differed from the blank-sky data sets by only $0.9\pm{1.8}$\%.
For the front illuminated S2 and S4 chips the
blank sky backgrounds were normalized downward by 5.9\% 
(uncertainty 2.1\%) and 14.9\% (uncertainty 2.0\%) respectively.
It is not surprising that the hard background surface brightness
in the blank-sky data might be higher than in our observations of DEM L50 and
DEM L152. The blank sky data sets are constructed from a large 
number of data sets and thus from data experiencing a higher (but
more normal) fraction of mild background flares than our observations. 

For the DEM L152 observation the blank sky background was normalized
down by 4.4\% (uncertainty $\pm{2.4}$\%).
For the S2 and S4 chips the
blank sky backgrounds were normalized downward by 2.0\% 
(uncertainty 3.0\%) and 16.6\% (uncertainty 2.8\%) respectively.

The magnitude of the difference in the mean source-free 
X-ray surface brightness in the $E=3$ -- 6 keV energy band 
for the S4 chip in both observations is somewhat surprising. This CCD chip
is known to suffer from significant flaws in its serial readout that leads
to artificial events being registered (visible as streaks in images) 
that can only be partially screened-out in software processing. We
speculate that this effect is normally exacerbated by the unwanted
particle events common in background flares, thus accounting for the
pronounced difference between the blank sky data for the S4 chip and
our observations.

We altered the coordinate system of the blank-sky data to match the
point of the appropriate observation. Thus our background subtraction
for image analysis consisted of subtracting an image created from the
blank-sky data from an image from the real observational data, scaled
by the ratio of total exposure times in the observation to the 
blank-sky data for the appropriate chip.

\begin{figure*}[!ht]
\epsscale{1.0}
\plotone{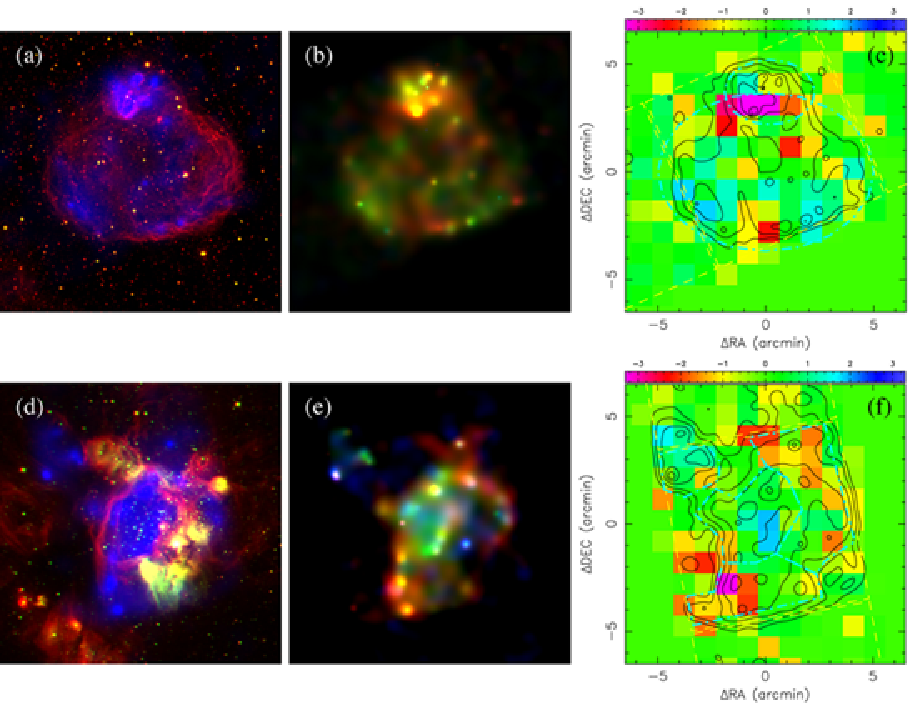}
  \caption{(a) A three-color optical/X-ray composite image of DEM L50
  using narrow-band optical \halpha+\nii~emission (red), \oiii~emission
  (green) and diffuse soft X-ray emission in the $E=0.3$--2.8 keV energy
  band (blue). The \halpha, \nii, and \oiii~image comes from \citet{oey02}.
  The intensity in each band is shown on a square-root intensity scale.
  (b) A three-color soft X-ray composite image of DEM L50 using
  the $E=0.3$--0.6 keV (red), $E=0.6$--1.6 keV (green) and $E=1.6$--2.8 
  keV (blue) energy bands. 
  The intensity in each band is shown on a linear intensity scale.
  (c) The statistical significance (in units of $\sigma$) 
  of the deviation of the local 
  X-ray spectral hardness
  ratio, assuming uniform absorption, from the mean bubble hardness ratio, as described in 
  \S~\ref{sec:spectra:regionselec}, overlaid with the contours of 
  $E=0.3$ -- 2.8 keV diffuse X-ray surface brightness.
  Panels (d), (e) and (f) are the equivalent of panels (a), (b) and (c),
  except that they show DEM L152. The \halpha, \nii, and \oiii~data is from \citet{smith99b}.
  The field of view in all panels is the same $13\arcmin \times 13\arcmin$
  region shown in Figs.~\ref{fig:img_deml50} and \ref{fig:img_deml152}.}
  \label{fig:rgb}
\end{figure*}

In Figures \ref{fig:img_deml50} and \ref{fig:img_deml152} we present 
adaptively smoothed X-ray images of DEM L50 and DEM L152 in a variety
of soft X-ray energy bands, along with optical \halpha+\nii~imaging
and \hi~column density maps. Fig.~\ref{fig:rgb} presents three-color
composite images of the superbubbles that combine narrow-band 
optical \halpha+\nii, \oiii~and soft X-ray emission.

Adaptive smoothing using the CIAO tool {\sc csmooth} 
\citep{ebeling06}  attempts
to differentially smooth an input image to achieve a relatively
uniform local S/N ratio, so that bright regions are smoothed with a smaller
smoothing kernel than low surface brightness regions. We initially
smoothed the diffuse image in the $E=0.3$ -- 2.8 keV energy band,
after point source removal but 
prior to background subtraction, with a target S/N ratio of 3 but with 
a maximum smoothing kernel equivalent to a Gaussian of FWHM$ = 40\arcsec$.
The smoothing map calculated in this step 
was then applied to all the background-subtracted soft X-ray diffuse 
images, so that any differences between images in different energy bands
are intrinsic and are not due to the application of different smoothing maps.

Given claims that {\sc csmooth} can generate spurious structure in adaptively
smoothed images \citep{diehl06}, we also investigated adaptively smoothed
images using the completely independent adaptive smoothing algorithm
implemented in the {\it XMM-Newton} 
data analysis software {\sc SAS} as {\sc asmooth}. This algorithm generated images very
similar to those produced by {\sc csmooth}, except that the stated 
statistical significance of individual features within the images differed
significantly between the two methods. We also 
generated images using uniform smoothing with a 2-dimensional
Gaussian mask of FWHM$=30\arcsec$ (not shown)
in order to conservatively verify the
existence of features seen in the adaptively smoothed images.

\subsection{DEM L50}
\label{sec:images:deml50}

The diffuse soft X-ray emission in DEM L50 is clearly confined within
the optical superbubble and SNR shells (Fig.~\ref{fig:img_deml50}b). 
The apparent X-ray surface brightness of the SNR N186 D \citep{oey02} on the 
northern edge of 
the roughly-elliptical superbubble is twice that of the larger superbubble.
If we ignore N186 D and concentrate solely on the superbubble we see that the
detected X-ray emission in the $E=0.3$ -- 2.8 keV energy band is 
relatively uniformly distributed, although with a slight tendency toward
being brighter on the south and eastern interior edges of the optical shell.

The \hi~column density along the line of sight to DEM L50 peaks in
the projected center of the bubble, however (see Fig.~\ref{fig:img_deml50}c), 
raising the possibility that the weak limb brightening is due to central
absorption rather than true enhancement of the X-ray emission at the shell
walls.

That at least some of the \hi~lies between us and the X-ray emitting 
plasma is clear from
the way the X-ray-bright southeast limb of the superbubble fills in
the gap in \hi~column density at that location (compare 
Fig.~\ref{fig:img_deml50}a and c). Based on the observed velocities of the \hi~gas, \citet{oey02} show that the \hi~associated
with DEM L50 is most likely the outer layer of the expanding shell of the bubble.

DEM L50 is most prominent in the $E=0.6$ 
-- 1.6 keV band. The SNR N186 D is prominent in both this energy band and the softer
$E=0.3$ -- 0.6 keV energy band, despite being associated with a
column density of $\nH \sim 10^{21} \pcmsq$ of \hi. This suggests
that the SNR may be spectrally distinct from the superbubble. Neither
the superbubble nor the SNR are apparent in the $E=1.6$ -- 2.8 keV 
energy band.

\begin{figure}[!ht]
\epsscale{1.0}
\plotone{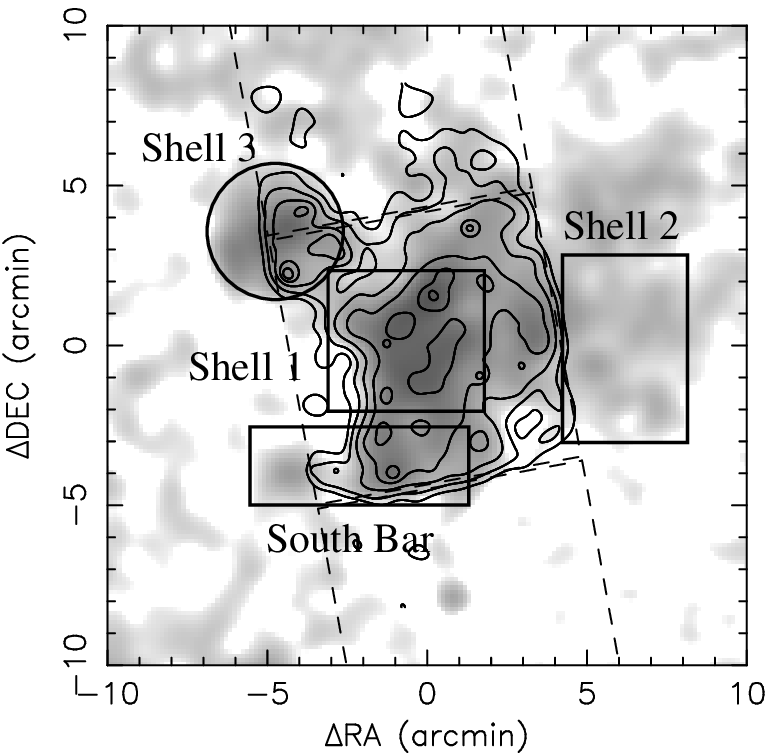}
  \caption{A comparison between the the {\it ROSAT} HRI (grey, 
  square root intensity scale) and 
  {\it Chandra} ACIS observations (contours, as in 
  Fig.~\ref{fig:img_deml152}a) 
  of N44.
  Dashed lines denote the edges of the {\it Chandra} ACIS 
  CCD chips. The features identified in \citet{chu93} are outlined
  and labeled.
  }
  \label{fig:hri_n44}
\end{figure}

\subsubsection{DEM L152}
\label{sec:images:deml152}

In contrast to the relative simplicity of DEM L50, the spatial structure of
DEM L152 (N44) in the X-ray, optical and \hi~is complex (see Fig.~\ref{fig:rgb}). 
The field of view of the {\it Chandra}
 observations (Fig.~\ref{fig:img_deml152}) 
cover the following
features identified in the {\it ROSAT} PSPC observations of \citet{chu93}:
Shell 1, the main superbubble, and parts of Shell 3, a SNR, and the South
Bar, in particular the suspected break-out region where Shell 1 may be venting
its contents through a break in its southern edge.

The relationship between the new {\it Chandra} observation and the features 
identified in prior X-ray observations is shown in Fig.~\ref{fig:hri_n44},
where the {\it Chandra} ACIS field of view and contours of diffuse X-ray 
surface brightness are overlaid on a smoothed {\it ROSAT} HRI image of N44. The image was taken from the 108.5 ks HRI observation RH600913, obtained from the HEASARC
data archive.

Shell 1 is the most prominent X-ray source in the $E=0.3$ -- 2.8 keV 
energy band {\it Chandra} image, and the soft diffuse X-ray emission is 
particularly bright around the inner edge of the optical shell.
This limb-brightening is not an artifact of adaptive smoothing, as it
is also clearly visible in images smoothed with a uniform FWHM$=30\arcsec$
Gaussian mask.

The limb-to-center brightness contrast is a factor of $\sim 2$ in the
$E=0.3$ -- 2.8 keV and $E=0.6$ -- 1.6 keV energy band images and is
slightly larger in the softer $E=0.3$ -- 0.6 keV energy band image 
(Fig.~\ref{fig:img_deml152}d). 

This central decrement in X-ray surface brightness cannot be due to 
absorption by intervening gas as both the X-ray-bright limbs and the
fainter center of Shell 1 lie within a region of lower \hi~column density
(Fig.~\ref{fig:img_deml152}c).

Shell 3 falls partly within the ACIS S3 and S2 chips, although its eastern
edge is not covered in the current observations. It has a comparable X-ray
surface brightness to Shell 1, even though the \hi~column density along
this line of sight ($\nH \sim 5 \times 10^{21} \pcmsq$) 
is approximately twice the value toward the
center of Shell 1.

Bright, diffuse, soft X-ray emission also extends to both the west and south
of Shell 1; the latter feature being known as the South Bar \citep{chu93}.

It is also noteworthy that both Shell 1 and Shell 3 are clearly detected
in the $E=1.6$ -- 2.8 keV energy band, in contrast to the complete lack
of emission in this energy band from DEM L50 and N186 D. The South Bar
lacks appreciable emission in this energy band, which is consistent
with the finding of \citet{magnier96} that the South Bar was spectrally
softer than Shell 1.

\section{Spectral analysis}
\label{sec:spectra}

\subsection{Spectral region selection}
\label{sec:spectra:regionselec}

\begin{deluxetable*}{llrlll}
\tabletypesize{\scriptsize}%
\tablecolumns{6}
\tablewidth{0pc}
\tablecaption{Count Rates
        \label{tab:rates}}
\tablehead{
\colhead{Target}    & \colhead{Region} & \colhead{Area (arcmin$^2$)}
  & \colhead{$R_{\rm tot}$}  & \colhead{$R_{\rm diff}$}
  & \colhead{$R_{\rm PS}$} 
    \\
\colhead{(1)} & \colhead{(2)} & \colhead{(3)} 
  & \colhead{(4)} & \colhead{(5)}
  & \colhead{(6)}
}
\startdata
DEM L50  & Bubble 
  &  46.9443 & $0.1830\pm{0.0031}$ & $0.1577\pm{0.0030}$ & $0.0252\pm{0.0009}$  \\
DEM L50  & SNR 
  &   9.8197 & $0.0776\pm{0.0018}$ & $0.0749\pm{0.0017}$ & $0.0027\pm{0.0003}$  \\
DEM L152  & Bubble 
  &  13.2994 & $0.2633\pm{0.0041}$ & $0.2626\pm{0.0041}$ & $0.0006\pm{0.0004}$  \\
DEM L152  & SNR 
  &   4.9459 & $0.0422\pm{0.0018}$ & $0.0422\pm{0.0018}$ & $0.0000\pm{0.0001}$ \\
DEM L152  & South Bar
  &  11.8618 & $0.1400\pm{0.0031}$ & $0.1362\pm{0.0031}$ & $0.0038\pm{0.0005}$  \\
DEM L152  & West 
  &  17.4595 & $0.1888\pm{0.0037}$ & $0.1855\pm{0.0036}$ & $0.0033\pm{0.0005}$ \\
\enddata
\end{deluxetable*}

As a first step to selecting regions for more detailed spectral analysis
we constructed three-color images of the two bubbles where the resultant
color is representative of the mean energy of the diffuse X-ray emission
(Fig.~\ref{fig:rgb}b \& e) within the soft X-ray band. In these panels
red represents the lowest energy photons ($E=0.3$ -- 0.6 keV), green
intermediate energy photons ($E=0.6$ -- 1.6 keV) and blue relatively harder 
emission in the $E=1.6$ -- 2.8 keV band.

As we shall later show, these figures somewhat exaggerate the spectral 
differences between the two objects and the degree of small-scale
spectral variation within each object. Nevertheless they do serve some 
purpose by more clearly illustrating some of the possible spatial variation in the
distribution of diffuse emission at different energies 
and help inform our
choice of regions over which to study the spectral properties of the 
diffuse emission.

In Fig.~\ref{fig:rgb}b the emission from the superbubble DEM L50 appears
slightly spectrally harder than the brighter emission from the SNR N186 D,
but there is only a marginal suggestion 
of spectral hardness variation within DEM L50 itself.
Based on this and the optical morphology of the bubble, we
choose to separate the X-ray data into two regions for
spectral fitting: An elliptical region encompassing SNR N186 D
and another elliptical region covering the superbubble but
excluding the SNR. These are the regions shown in
Fig.~\ref{fig:img_deml50}a. In order to examine spectral variations on smaller scales, we further subdivided DEM L50 into a Center and Limb region when performing spectral fits (see \S~\ref{sec:spectra:data}). Note that only the parts of those ellipses within
the S3 chip are actually included within the spectral extraction regions.

The soft X-ray emission in the vicinity of DEM L152 shows more evident
spectral variation than does DEM L50 (Figure~\ref{fig:rgb}e). Based on this and the optical
morphology of the N44 nebula we chose spectral extraction regions
similar to those used earlier by \citet{chu93} and \citet{magnier96}: Shell
1 (the main bubble), Shell 3 (the SNR), and the South Bar, with most of the remaining emission associated
with a new region we simply name West (see Fig.~\ref{fig:img_deml152}a).

Diffuse X-ray count rates within the spectral
extraction regions of both sets of observations are given in 
Table~\ref{tab:rates}. Column 3 gives the area of each region. Columns 4-6 show the background-subtracted ACIS S3 count rates in the $E=0.3-2.8$ keV energy band within each specified region, where $R_{\rm tot}$ gives the count rate for the full X-ray emission, $R_{\rm diff}$ gives the estimated diffuse emission count rate, and $R_{\rm PS}$ gives the total point source count rate. Emission from point-like X-ray
sources accounts for $\sim 14$\% of the total soft X-ray emission within
the boundaries of the DEM L50 bubble, and less than $1$\% of the emission
within the boundary of the DEM L152 Shell 1.

Given {\it Chandra}'s high spatial resolution and sensitivity we should be able to determine
whether the spectral properties of the diffuse emission vary on 
spatial scales smaller than the entire shell or bubbles. In Figures~\ref{fig:rgb}b \& e we show energy-color-coded images of the
diffuse emission around DEM L50 and DEM L152. 
Such images can be prone to smoothing-related artifacts, so we also mapped the
hardness ratios in $1\arcmin \times 1\arcmin$ pixels. These large pixels
were required in order to minimize the statistical uncertainty in the hardness
ratio in each pixel.
The spectral hardness ratio, which will be discussed in more detail in \S~\ref{sec:spectra:data}, is defined as $Q=(H-S)/(H+S)$, where $H$ and $S$ are the
count rates in the specified hard ($E = 0.6-1.6$ keV) and soft ($E = 0.3-0.6$ keV) energy bands respectively. Thus
higher values of $Q$ correspond to a harder spectrum. 
Images of the hardness ratio itself tend to be difficult to interpret, as
they can be dominated by noise in the lower signal-to-noise regions. For this
reason we constructed maps of the statistical significance of the deviation
of the local hardness ratio from the average 
hardness ratio $\overline{Q}$ over each 
bubble. For a given
pixel with hardness ratio $Q_{\rm pix}$ and uncertainty $\sigma_{\rm Q, pix}$, Figures~\ref{fig:rgb}c \& f 
show $(Q_{\rm pix} - \overline{Q})/\sigma_{\rm Q, pix}$. In other words, the value in each pixel corresponds to the significance of the deviation in Gaussian $\sigma$. While Figures~\ref{fig:rgb}c  \& f hint at some of the intrinsic large-scale variations across the superbubbles, variations in column density and the resultant absorption of soft X-ray photons have a substantial effect on the observed hardness ratios. Figures~\ref{fig:rgb}c \& f should only be used to identify regions that appear spectrally distinct from one another. 

Although some spatial variation in the spectral
properties of the diffuse X-ray emission is present in both objects,
it is not particularly strong, and emission in the $E=0.6$ -- 1.6 keV
energy band dominates both sets of observations. Hardness ratio variations are
present on large scales, for example between the bubbles and the
nearby supernova remnant, or between Shell 1 and the Southern Bar 
region of DEM L152 . On smaller angular scales ($\sim 1\arcmin$)
we find that any spatial variations in spectral hardness are of
marginal statistical significance in our current X-ray data.

\subsection{Spectral fitting}
\label{sec:spectra:data}
For the spectral fits, we removed background flares from both the S3 and S2 chips with the {\sc lc\_clean} routine in CIAO. We then used the re-normalized ACIS stowed background files to subtract the approximately constant particle background from both chips. To re-normalize the background files, we compared the relative count rates of the stowed background and our observations between 8 keV and 9.4 keV and re-normalized the particle background to match the count rate of our data. The $8-9.4$ keV energy range was chosen to exclude real X-ray photon events and a line feature above 9.4 keV. 

We used the {\sc specextract} script in CIAO to extract spectra from the specified regions and subtract the particle background. To account for the X-ray background, we fit the extracted spectra from the S2 chip,  following \citet{kuntz00} and \citet{henley07}, and used the resulting fits as a component of our subsequent models to the S3 data. A thermal plasma in collisional ionization equilibrium fits the local X-ray background, while two additional thermal plasmas in equilibrium are used for the Galactic background, and the cumulative emission from extragalactic sources is fit by a power law. Our best fits to the S2 background are shown in Table~\ref{tab:fits2}. The power law index was fixed to the value found by \citet{chen97}, and the column densities were obtained from HEASARC from \citet{kalberla05}. Temperatures were left as a free parameter. The Local Hot Bubble temperature is given in Column 3, the Galactic \hi~column density is shown in Column 4, the temperatures of the two thermal components of the Galactic background are given in Columns 5 and 6, and the extragalactic power law index is shown in Column 7. Normalizations were re-scaled to account for the relative difference in area between the S3 extracted regions and the S2 chip. For the sake of comparison, we include two fits to the S2 background for DEM L50; the ``Low S2" fit was a local minimum in $\chi^2$ before the global best fit was found. The effect of varying the background fit can be seen by comparing the DEM L50 models with the ``High S2" and ``Low S2" background fits.

\begin{figure*}[!ht]
\epsscale{1.0}
\plottwo{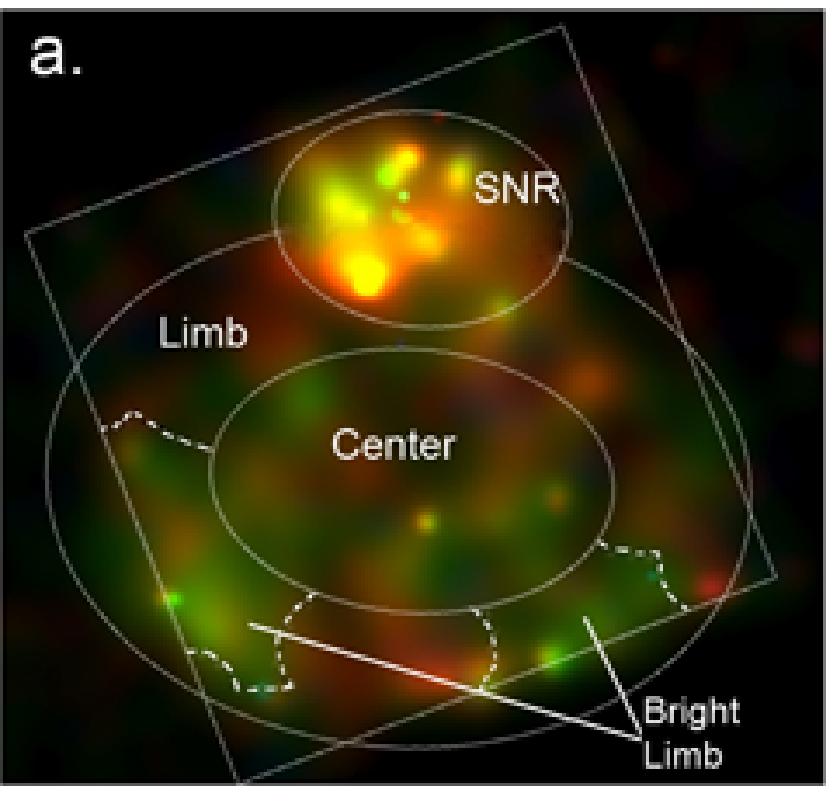}{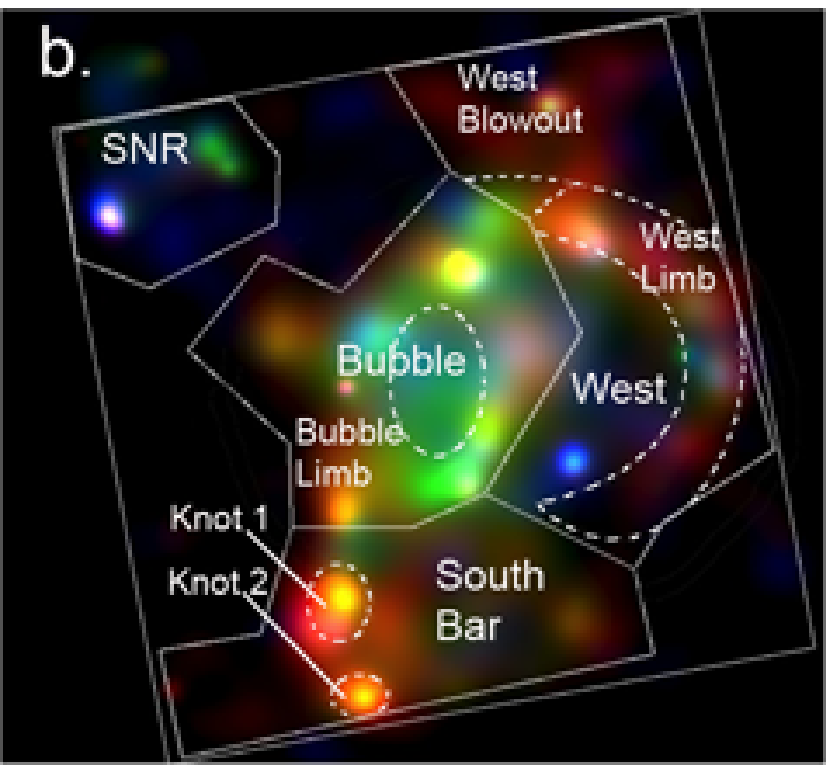}
  \caption{Colors and smoothing as in Fig.~\ref{fig:rgb}b. a. Extracted regions (solid lines) and subregions (dashed lines) for DEM L50. b. Extracted regions (solid lines) and subregions (dashed lines) for DEM L152.
  }
  \label{fig:ext_regions}
\end{figure*}

All extracted regions are shown in Figure~\ref{fig:ext_regions}. In addition to the regions in Figs.~\ref{fig:img_deml50}a and~\ref{fig:img_deml152}a, we extracted spectra for smaller subregions. We divided DEM L50 into Limb and Center regions a separate spectrum for the bright southern portion of the Limb. For DEM L152, we extracted additional spectra for the Bubble (that is, Shell 1) Limb, West Limb, and West Blowout regions as well as for the two knots shown in Fig.~\ref{fig:ext_regions}. While some clumps seen in the smoothed image of DEM L152 are artifacts of the smoothing algorithm, these two knots are statistically significant (see Fig.~\ref{fig:rgb}f) and appear when different smoothing algorithms are used. 

We fit each superbubble region and subregion with one- and two-temperature plasmas in collisional ionization equilibrium, using the XSPEC Astrophysical Plasma Emission Code (APEC) models \citep{smith01} with variable abundances (Tables~\ref{tab:fit50} and \ref{tab:fit152}). We examined an additional, non-equilibrium ionization model for DEM L50 (Table~\ref{tab:fit50ne}). In all models, He, C, N, Al, and Ni abundances were fixed to the solar values. We varied the $\alpha$ elements O, Ne, Mg, Si, S, Ar, and Ca in proportion to one another, and Fe was left as a free parameter. Although absolute abundances are not well constrained, the $\alpha$/Fe ratio is more reliable. To ensure that the fits were not local minima, we looked at possible values of \nH, temperature, and abundances spanning one to two orders of magnitude with the XSPEC command {\sc steppar}. As a further check on the validity of our fits and to establish error bars for each variable, we ran the {\sc error} command on the free parameters. 

To determine fluxes, we ran the XSPEC command {\sc flux} on each model and determined the absorption-corrected fluxes by setting \nH~to zero. We adopted the distance of 50 kpc to the LMC when converting absorption-corrected fluxes to luminosities. Observed fluxes and the calculated luminosities are shown in Tables~\ref{tab:fluxobs50}-\ref{tab:lumin152}, which we discuss below.

We also calculated hardness ratios and median energies for the diffuse emission from each region, shown in Table~\ref{tab:hardness}; the hardness ratio $Q=(H-S)/(H+S)$, where $H$ and $S$ are the
luminosities in the hard ($E=0.6$ -- 1.6 keV) energy band and soft ($E=0.3$ -- 0.6 keV) energy band. Absorption by intervening \hi~results in an apparently harder spectrum, and Figs.~\ref{fig:img_deml50}c \&~\ref{fig:img_deml152}c show that there is significant variation in the \hi~column density across these objects. The column density was a free parameter in our fits, and the resulting luminosities and hardness ratios account for \nH~variations between regions. $Q$ obtained using the count rate (i.e. not absorption-corrected) is shown for comparison in Column 4 of Table~\ref{tab:hardness}. The influence of the column density can be seen by comparing the luminosity-determined and count rate-determined hardness ratios for the SNR 0523-679 \citep{chu93} in DEM L152. While the SNR appears to have the hardest spectrum from the count rates, its higher absorption column density masks a much softer spectrum. 

\citet{hong04} demonstrate that the median energy is a more reliable indicator of X-ray spectral variations than the hardness ratio is. As a result, we include the median energy of each region in Table~\ref{tab:hardness}. The median energies generally agree with the hardness ratios. An exception is the DEM L152 SNR, which has a higher median energy due to the \hi~absorption discussed above. 

\begin{deluxetable*}{lllllll}
\tabletypesize{\scriptsize}%
\tablecolumns{8}
\tablewidth{0pc}
\tablecaption{Parameters for Background Fits
        \label{tab:fits2}}
\tablehead{
\colhead{Bubble}    & \colhead{Background Fit} & \colhead{$kT_{\rm LHB}$ (keV)} & \colhead{\nH\tablenotemark{a} ($10^{22}$ cm$^{-2}$})
   & \colhead{$kT_{1}$ (keV)}
  & \colhead{$kT_{2}$ (keV)}  & \colhead{Photon Index\tablenotemark{b}}
  \\
\colhead{(1)} & \colhead{(2)} & \colhead{(3)}
  & \colhead{(4)} & \colhead{(5)}
  & \colhead{(6)} & \colhead{(7)}
  }
\startdata

DEM L50  & High S2 Fit 
  &   1.08  & 0.132 & $0.05$ & $0.22$ 
  &  $1.46$  \\
DEM L50  & Low S2 Fit 
  &  0.08  & 0.132 & $0.10$ & $0.24$ 
  &  $1.46$ \\
DEM L152  & L152 S2 Fit 
  &  0.08 & 0.228 & $0.06$ & $0.25$ 
  &  $1.46$ \\

\enddata
\tablenotetext{a}{From \citet{kalberla05}, provided by HEASARC.}
\tablenotetext{b}{From \citet{chen97}.}
\end{deluxetable*}

\begin{turnpage}
\begin{deluxetable*}{lllllllllllll}
\tabletypesize{\scriptsize}%
\tablecolumns{13}
\tablewidth{0pc}
\tablecaption{Fit Parameters for DEM L50
        \label{tab:fit50}}
\tablehead{
\colhead{Model}    & \colhead{Bkgd Fit\tablenotemark{a}} & \colhead{\nH ($10^{22}$ cm$^{-2}$)}
  & \colhead{$kT_{1}$ (keV)}  & \colhead{norm$_{1}$}
  & \colhead{$kT_{2}$ (keV)}  & \colhead{norm$_{2}$}
  & \colhead{eff $kT$ (keV)}
  & \colhead{$\alpha$/O$_{\sun}$} & \colhead{Fe/Fe$_{\sun}$}
  & \colhead{$\frac{\alpha/Fe}{(\alpha/Fe_{\sun})}$}
  & \colhead{red. $\chi^{2}$} & \colhead{D.O.F}
  \\
\colhead{(1)} & \colhead{(2)} & \colhead{(3)} 
  & \colhead{(4)} & \colhead{(5)}
  & \colhead{(6)} & \colhead{(7)}
  & \colhead{(8)} & \colhead{(9)} & \colhead{(10)}
  & \colhead {(11)} & \colhead{(12)} & \colhead{(13)}
}
\startdata
{\bf Center 1T} & {\bf High S2}
  & ${\bf  0.38^{+0.03}_{-0.04}}$ & ${\bf 0.15^{+0.02}_{-0.01}}$  & {\bf 3.5E-03 }& ${\bf ...}$
  & ${\bf ...}$ & {\bf 0.15 }& ${\bf 0.88^{+0.87}_{-0.35} }$ & ${\bf 6.00^{+12.87}_{-3.42}}$ 
  & ${\bf 0.15^{+0.05}_{-0.04}}$ & {\bf 1.743} & {\bf 211} \\[+0.1cm]
 Center 1T solar & High S2
  & 0.31 $\pm$ 0.05 & 0.19 $\pm$ 0.01 & 1.3E-03 & $...$
  & $...$ & 0.19 & 1.00 & 1.00 & 1.00 & 1.943 & 213 \\[+0.1cm]
 Center 2T & High S2
  & $0.40^{+0.08}_{-0.06}$  & $0.11^{+0.80}_{-0.11}$ & 4.7E-03 &  0.17 $\pm$ 0.03
  & 1.6E-03 & 0.13 & $1.10^{+1.18}_{-0.60}$ & $6.03^{+10.61}_{-3.35}$
  & $0.18^{+0.14}_{-0.05}$ & 1.757 & 209 \\[+0.1cm]
  Center 2T solar & High S2
  & $0.32^{+0.06}_{-0.09}$ & $0.11^{+0.04}_{-0.02}$ & 3.1E-03 & $0.23^{+0.06}_{-0.02}$ 
  & 5.9E-04 & 0.13 & 1.00 & 1.00 & 1.00 & 1.835 & 211\\[+0.1cm]
 Limb 1T & High S2
  & 0.29 $\pm$ 0.05 & $0.18^{+0.02}_{-0.04}$ & 2.7E-03 & $...$
  & $...$ & 0.18 & $0.54^{+0.29}_{-0.17}$ & $1.93^{+8.04}_{-0.94}$
  & $0.28^{+0.10}_{-0.06}$ & 1.744 & 302 \\[+0.1cm]
 {\bf Limb 2T} & {\bf High S2}
  & ${\bf 0.32^{+0.03}_{-0.02} }$& ${\bf 0.08^{+0.20}_{-0.08}  }$& {\bf 1.4E-02 }& ${\bf 0.22^{+0.07}_{-0.02}}$ 
  & {\bf 7.3E-04 }& {\bf 0.09 }& ${\bf 1.25^{+0.62}_{-0.36}}$  &${\bf 2.78^{+1.51}_{-1.37}}$
  & ${\bf 0.45^{+0.27}_{-0.17} }$& {\bf 1.718 }&{\bf  300} \\[+0.1cm]
 Limb 2T solar & High S2
  & 0.21 $\pm$ 0.04 & $0.11^{+0.04}_{-0.02}$  & 1.4E-03 & $0.27^{+0.03}_{-0.04}$ 
  & 2.1E-04 & 0.13 & 1.00 & 1.00 & 1.00 & 1.749 & 302 \\[+0.1cm]
Limb 1T & Low S2
  & 0.36 $\pm$ 0.06 & $0.18^{+0.01}_{-0.03}$ & 2.6E-03 & $...$
  & $...$ & 0.18 & $0.95^{+2.70}_{-0.39}$  & $3.38^{+13.92}_{-1.95}$ 
  & $0.28^{+0.12}_{-0.08}$ & 1.793 & 302 \\[+0.1cm]
 Limb 2T & Low S2
  & $0.34^{+0.07}_{-0.11}$ & $0.14^{+0.26}_{-0.14}$ & 1.4E-03 & $0.21^{+0.16}_{-0.11}$ 
  & 7.1E-04 & 0.16 & $1.37^{+2.50}_{-0.71}$ & $3.58^{+9.85}_{-2.05}$ 
  & $0.38^{+0.28}_{-0.16}$ & 1.799 & 300 \\[+0.1cm]
 {\bf Bright Limb 1T } & {\bf High S2}
  & ${\bf 0.37^{+0.07}_{-0.08} }$ & ${\bf 0.19^{+0.02}_{-0.03} }$ & {\bf 1.9E-03 } & ${\bf ...}$
  & ${\bf ...}$ & {\bf 0.19} & ${\bf 0.56^{+0.95}_{-0.25}}$  &${\bf 1.07^{+7.40}_{-1.23}}$ 
  & ${\bf 0.53^{+0.27}_{-0.18}}$ &{\bf  1.516} & {\bf 137} \\[+0.1cm]
 Bright Limb 2T & High S2
  & 0.37 $\pm$ 0.07 & $0.11^{+0.15}_{-0.11}$ & 1.8E-03 & $0.22^{+0.11}_{-0.05}$ 
  & 8.3E-04 & 0.14 & $0.80^{+2.35}_{-0.46}$  & $1.21^{+2.72}_{-0.72}$ 
  & $0.66^{+0.44}_{-0.37}$ & 1.532 & 135 \\[+0.1cm]
 {\bf SNR 2T} & {\bf High S2}
  &${\bf  0.09 \pm 0.02 }$& ${\bf 0.19 \pm 0.01 }$& {\bf 4.7E-04 }&${\bf  0.72^{+0.09}_{-0.05}}$
  & {\bf 2.6E-05 }& {\bf 0.22} & ${\bf 0.37^{+0.23}_{-0.11}}$ &${\bf 1.66^{+2.21}_{-0.81}}$ 
  & ${\bf 0.22^{+0.13}_{-0.06} }$&{\bf  2.114} & {\bf 145} \\[+0.1cm]
 SNR 2T solar & High S2
  & $0.20^{+0.05}_{-0.03}$ & $0.08^{+0.01}_{-0.08}$ & 6.2E-03 & 0.29 $\pm$ 0.02
  & 2.8E-04 & 0.09 & 1.00 & 1.00 & 1.00 & 2.173 & 147 \\[+0.1cm]
 SNR 2T & Low S2
  & $0.06^{+0.05}_{-0.04}$ & $0.19^{+0.02}_{-0.01}$ &4.3E-04 & 0.73 $\pm$ 0.09
  & 3.2E-05 & 0.23 & $0.27^{+0.12}_{-0.06}$ & $1.26^{+1.50}_{-0.54}$
  & $0.22^{+0.11}_{-0.06}$ & 2.150 & 145 \\[+0.1cm]
 SNR 2T solar & Low S2
  & $0.25^{+0.04}_{-0.06}$ & $0.08^{+0.01}_{-0.08}$ & 9.9E-03 & $0.27^{+0.02}_{-0.01}$ 
  & 3.9E-04 & 0.09 & 1.00 & 1.00 & 1.00 & 2.255 & 148 \\
 
\enddata
\tablecomments{
  Errors show the 90\% confidence interval when one parameter is varied.
}
\tablenotetext{a}{The Low S2 background gave poor fits to the Center region which are not included.} 
\end{deluxetable*}
\end{turnpage}

\begin{turnpage}
\begin{deluxetable*}{llllllllllll}
\tabletypesize{\scriptsize}%
\tablecolumns{12}
\tablewidth{0pc}
\tablecaption{DEM L50 Non-Equilibrium Fits
        \label{tab:fit50ne}}
\tablehead{
\colhead{Model}    & \colhead{Bkgd Fit\tablenotemark{a}} & \colhead{\nH ($10^{22}$ cm$^{-2}$)}
  & \colhead{$kT_{1}$ (keV)}  & \colhead{norm$_{1}$}
  & \colhead{$\tau$ (s cm$^{-3}$)}  & \colhead{Ion. Time. (yr)}
  & \colhead{$\alpha$/O$_{\sun}$} & \colhead{Fe/Fe$_{\sun}$}
  & \colhead{$\frac{\alpha/Fe}{(\alpha/Fe_{\sun})}$}
  & \colhead{red. $\chi{^2}$} & \colhead{D.O.F.}
  \\
\colhead{(1)} & \colhead{(2)} & \colhead{(3)} 
  & \colhead{(4)} & \colhead{(5)}
  & \colhead{(6)} & \colhead{(7)}
  & \colhead{(8)} & \colhead{(9)} & \colhead{(10)}
  & \colhead{(11)} & \colhead{(12)}
}
\startdata
Center NEI & High S2
 & $0.31^{+0.05}_{-0.08}$ & $0.28^{+0.07}_{-0.06}$ & $5.1^{+12.0}_{-3.3}$E-04 
 & 5.57E10 $\pm$ 4.68E10 & 22,000 & $0.76^{+0.36}_{-0.24} $ & $0.74^{+0.36}_{-0.26}$
 & $1.04^{+0.27}_{-0.20}$ & 1.838 & 210 \\[+0.1cm]
Limb NEI & High S2
 & $0.26^{+0.04}_{-0.03}$  & $0.34^{+0.11}_{-0.05}$ & $3.2^{+6.3}_{-2.3}$E-04
 & 3.31E10 $\pm$ 2.84E10 & 46,000 & $0.83^{+0.46}_{-0.24}$  & $1.10^{0.62}_{-0.34}$ 
 & $0.76^{+0.16}_{-0.12}$ & 1.718 & 301 \\[+0.1cm]
Limb NEI & Low S2
 & $0.34^{+0.09}_{-0.08}$ & $0.28^{+0.11}_{-0.07}$ & $5.5^{+18.7}_{-4.4}$E-04
 & 6.26E10 $\pm$ 4.69E10 & 65,000 & $1.28^{+1.48}_{-0.49}$ & $1.51^{+0.87}_{-0.60}$
 & $0.85^{+0.19}_{-0.15}$ & 1.718 & 301 \\[+0.1cm]
Bright Limb NEI & High S2
 & $0.34^{+0.12}_{-0.10}$ & $0.32^{+0.14}_{-0.10}$ & $3.8^{+15.8}_{-1.8}$E-04
 & 8.33E10 $\pm$ 2.88E10 & 53,000\tablenotemark{b} & $0.70^{+0.69}_{-0.36}$ & $0.58^{+0.55}_{-0.27}$
 & $1.21^{+0.38}_{-0.27}$ & 1.543 & 136 \\[+0.1cm]
SNR NEI & High S2
 & $0.11^{+0.05}_{-0.03}$ & $0.58^{+0.23}_{-0.08}$ & $1.2^{+0.04}_{-0.06}$E-04
 & 2.60E10 $\pm$ 6.39E9 & 11,000 & $0.29^{+0.10}_{-0.09}$ & $0.45^{+0.14}_{-0.08}$
 & $0.64^{+0.15}_{-0.12}$ & 2.265 & 146 \\

\enddata
\tablecomments{
 Errors show the 90\% confidence interval when one parameter is varied. $\tau$ errors are estimates only. 
}
\tablenotetext{a}{The Low S2 background gave poor fits to the Center region which are not included.} 
\tablenotetext{b}{assuming $n_e~\sim0.05$ cm$^{-3}$.}
\end{deluxetable*}
\end{turnpage}

\begin{turnpage}
\begin{deluxetable*}{llllllllllll}
\tabletypesize{\scriptsize}%
\tablecolumns{12}
\tablewidth{0pc}
\tablecaption{Fit Parameters for DEM L152
        \label{tab:fit152}}
\tablehead{
\colhead{Region and Model}    & \colhead{\nH ($10^{22}$ cm$^{-2}$)}
  & \colhead{$kT_{1}$ (keV)}  & \colhead{norm$_{1}$}
  & \colhead{$kT_{2}$ (keV)}  & \colhead{norm$_{2}$}
  & \colhead{eff $kT$ (keV)}
  & \colhead{$\alpha$/O$_{\sun}$} & \colhead{Fe/Fe$_{\sun}$}
  & \colhead{$\frac{\alpha/Fe}{(\alpha/Fe_{\sun})}$}
  & \colhead{red. $\chi^{2}$} & \colhead {D.O.F}
  \\
\colhead{(1)} & \colhead{(2)} & \colhead{(3)} 
  & \colhead{(4)} & \colhead{(5)}
  & \colhead{(6)} & \colhead{(7)}
  & \colhead{(8)}  & \colhead{(9)}
  & \colhead{(10)} & \colhead {(11)} & \colhead {(12)}
}
\startdata
{\bf Bubble 1T} 
  &  ${\bf 0.27 \pm 0.02 }$&${\bf  0.37 \pm 0.02 }$& {\bf 2.5E-03 }&${\bf  ...}$ 
  & ${\bf  ...}$ &{\bf  0.37 }& ${\bf 0.68^{+0.61}_{-0.26} }$&${\bf  0.21^{+0.16}_{-0.07}}$
  & ${\bf 3.27^{+0.58}_{-0.43}}$ & {\bf 1.360 }& {\bf 151} \\[+0.1cm]
Bubble 2T 
  &   $0.29^{+0.04}_{-0.02}$ & $0.17^{+0.14}_{-0.07}$ & 5.5E-04 & $0.37^{+0.05}_{-0.02}$
  &  2.2E-03 & 0.33 & $0.76^{+0.78}_{-0.27}$ & $0.26^{+0.12}_{-0.10}$ 
  & $2.95^{+0.65}_{-0.47}$ & 1.352 & 149 \\[+0.1cm]
{\bf Bubble Limb 1T}
  & ${\bf 0.28 \pm 0.03}$ & ${\bf 0.33 \pm 0.02}$ & {\bf 3.3E-03 }& ${\bf ...}$
  & ${\bf ...}$ &{\bf  0.33 }& ${\bf 0.48^{+0.38}_{-0.17} }$& ${\bf 0.18 \pm 0.05}$
  & ${\bf 2.65^{+0.59}_{-0.43} }$& {\bf 1.794 }& {\bf 138} \\[+0.1cm]
 Bubble Limb 2T \nH~fixed
  & 0.28 & $0.14^{+0.10}_{-0.04}$ & 4.7E-04 & $0.35^{+0.02}_{-0.01}$
  & 1.7E-03 & 0.30 & $0.88^{+0.92}_{-0.35}$  & $0.32^{+0.39}_{-0.15}$
  & $2.76^{+0.55}_{-0.38}$ & 1.782 & 137 \\[+0.1cm]
West 1T  
  &  $0.27^{+0.03}_{-0.02}$ & 0.30 $\pm$ 0.02 & 3.6E-03 & $...$ 
  &  $...$ & 0.30 & $0.36^{+0.24}_{-0.11}$ & $0.10^{+0.07}_{-0.04}$
  & $3.42^{+1.34}_{-0.77}$ & 1.088 & 154 \\[+0.1cm]
 {\bf West 2T}
  & ${\bf 0.28^{+0.04}_{-0.03} }$& ${\bf 0.25^{+0.04}_{-0.03} }$&{\bf  2.5E-03 }&${\bf  0.79^{+0.15}_{-0.21}}$ 
  & {\bf 2.9E-04 }& {\bf 0.31 }& ${\bf 0.50^{+0.43}_{-0.18}}$ &${\bf  0.29^{+0.29}_{-0.15}}$
  & ${\bf 1.74^{+0.97}_{-0.58}}$ & {\bf 1.027} & {\bf 152} \\[+0.1cm]
 West Limb 1T\tablenotemark{c}%
  & $0.24^{+0.06}_{-0.03}$ & $0.32^{+0.04}_{-0.05}$ & 7.3E-04 & $...$
  & $...$ & 0.32 & $0.49^{+16.89}_{-0.29}$ & $0.21^{+2.28}_{-0.09}$
  & $2.31^{+1.21}_{-1.89}$ & 1.387 & 78 \\[+0.1cm]
 West Limb 2T \nH~fixed\tablenotemark{c}
  & 0.28 & $0.21^{+0.03}_{-0.02}$ & 6.6E-04 & $0.83^{+0.09}_{-0.14}$
  & 5.7E-05 & 0.26 & $0.68^{+6.12}_{-0.36}$ & $1.22^{+2.33}_{-0.75}$ 
  & $0.56^{+0.35}_{-0.22}$ & 1.355 & 77 \\[+0.1cm]
 {\bf West Blowout 1T \nH~fixed}
& {\bf  0.28 }& ${\bf 0.25 \pm 0.02}$ &{\bf  2.0E-04} & ${\bf ...}$
  & ${\bf ... }$& {\bf 0.25 }& ${\bf 2.34\tablenotemark{b}}$%
  &${\bf  0.32^{+2.56}_{-0.32}}$
  & ${\bf 7.39_{-4.22} }$& {\bf 1.143 }& {\bf 77} \\[+0.1cm]
 West Blowout 2T \nH~fixed
  & 0.28 & $0.14^{+4.30}_{-0.14}$ & 2.5E-05 & $0.27^{+0.28}_{-0.04}$
  & 3.7E-05 & 0.22 & 10.10\tablenotemark{b}%
  & $1.64^{+374.11}_{-1.64}$
  & $6.08_{-3.23}$ & 1.156 & 75 \\[+0.1cm]
 South Bar 1T\tablenotemark{c}%
  & 0.23 $\pm$ 0.02 & $0.35^{+0.03}_{-0.02}$ & 1.0E-03 & $...$ & $...$
  & 0.35 & $0.75^{+2.71}_{-0.35}$ & $0.21^{+0.70}_{0.08}$
  & $3.56^{+1.35}_{-0.74}$ & 1.054 & 128 \\[+0.1cm]
 South Bar 2T\tablenotemark{c}%
  & 0.21$\pm$ 0.02 & $0.29^{+0.04}_{-0.05}$ & 4.4E-05 & $0.86^{+0.06}_{-0.07}$
  & 1.5E-05 & 0.43 & $10.00^{+21.02}_{-7.23}$ & 6.84\tablenotemark{b}%
  & $1.46^{+1.75}_{-0.46}$ & 1.034 & 126 \\[+0.1cm]
 {\bf SNR 1T}
  & ${\bf 0.34^{+0.06}_{-0.04}}$ & ${\bf 0.38^{+0.07}_{-0.06}}$ & {\bf 9.3E-04} & ${\bf ...}$
  & ${\bf ...}$ & {\bf 0.38 }& ${\bf 0.36^{+0.96}_{-0.21}}$ &${\bf  0.04^{+0.05}_{-0.04}}$
  & ${\bf 10.24^{+24.59}_{-5.15}}$ &{\bf  1.176} &{\bf  67} \\[+0.1cm]
 SNR 2T
  & $0.35^{+0.05}_{-0.04}$ & $0.35^{+0.08}_{-0.23}$ & 7.4E-04 & 0.86\tablenotemark{b}%
  & 6.3E-05 & 0.39 & $0.44^{+0.64}_{-0.27}$ & $0.05^{+0.06}_{-0.05}$
  & $8.18^{+51.18}_{-6.99}$ & 1.198 & 65 \\[+0.1cm]
 Knot 1 1T solar
  & 0.22 & 0.30 $\pm$ 0.03 &  8.2E-05 & $...$
  & $...$ & 0.30 & 1.00 & 1.00 & 1.00
  & 1.471 & 17 \\[+0.1 cm]
 Knot 1 1T free abun
  & 0.22 & 0.30 $\pm$ 0.03 & 7.3E-05 & $...$
  & $...$ & 0.30 & 1.40$_{-1.07}$\tablenotemark{b} %
  & $0.53^{+9.87}_{-0.33}$ & $2.63^{+4.67}_{-1.20}$
  & 1.253 & 15 \\[+0.1 cm]
 Knot 2 1T solar
  & 0.22 & $0.30^{+0.07}_{-0.05}$ & 3.5E-05 & $...$
  & $...$ & 0.30 & 1.00 & 1.00 & 1.00 & 1.288 & 8\\

\enddata
\tablecomments{
   Errors show the 90\% confidence interval when one parameter is varied. 
 }
 \tablenotetext{b}{Errors are unconstrained.}
 \tablenotetext{c}{It is not clear which fit is statistically better. The {\it p}-value for a 2T fit is $\sim0.1$.} 

\end{deluxetable*}
\end{turnpage}

\begin{deluxetable*}{llllll}
\tabletypesize{\scriptsize}%
\tablecolumns{6}
\tablewidth{0pc}
\tablecaption{Observed Fluxes for DEM L50
        \label{tab:fluxobs50}}
\tablehead{
\colhead{Model}    & \colhead{0.3-2.0 keV}
  & \colhead{0.5-2.0 keV}  & \colhead{0.3-8.0 keV}
  & \colhead{0.5-8.0 keV}  & \colhead{2.0-8.0 keV}
  \\
\colhead{(1)} & \colhead{(2)} & \colhead{(3)} 
  & \colhead{(4)} & \colhead{(5)}
  & \colhead{(6)} 
}
\startdata
Center 1T
 & 0.24 &  0.23 & 0.28 & 0.26 & 0.032 \\
 Center 1T solar
 & 0.24 & 0.22 & 0.27 & 0.25 & 0.032 \\
 Center 2T
 & 0.24 & 0.23 & 0.28 & 0.26 & 0.031 \\
 Center 2T solar
 & 0.25 & 0.23 & 0.28 & 0.26 & 0.032 \\
 {\bf Center NEI}
 & {\bf 0.25} & {\bf 0.22} & {\bf 0.28} & {\bf 0.26} & {\bf 0.032} \\
 Limb 1T
 & 0.39 & 0.36 & 0.45 & 0.42 & 0.057 \\
 Limb 2T
 & 0.41 & 0.36 & 0.47 & 0.42 & 0.057 \\
 Limb 2T solar
 & 0.43 & 0.36 & 0.48 & 0.42 & 0.058 \\
 Limb 1T $(Low S2)$
 & 0.48 & 0.36 & 0.54 & 0.43 & 0.063 \\
 Limb 2T $(Low S2)$
 & 0.55 & 0.36 & 0.55 & 0.43 & 0.064 \\
 {\bf Limb NEI}
  & {\bf 0.41} & {\bf 0.36} & {\bf 0.46} & {\bf 0.42} & {\bf 0.057} \\
 Limb NEI $(Low S2)$
 & 0.48 & 0.36 & 0.55 & 0.43 & 0.064 \\
 Bright Limb 1T
 & 0.16 & 0.15 & 0.17 & 0.17 & 0.015 \\
 Bright Limb 2T
 & 0.16 & 0.15 & 0.18 & 0.17 & 0.015 \\
 {\bf Bright Limb NEI}
 & {\bf 0.16} & {\bf 0.15} & {\bf 0.18} & {\bf 0.17} & {\bf 0.015} \\
 SNR 2T
 & 0.37 & 0.26 & 0.39 & 0.28 & 0.021 \\
 SNR 2T solar
 & 0.34 & 0.26 & 0.36 & 0.28 & 0.020\\
 SNR 2T $(Low S2)$
 & 0.36 & 0.26 & 0.39 & 0.29 & 0.023 \\
 SNR 2T solar $(Low S2)$
 & 0.35 & 0.26 & 0.37 & 0.28 & 0.022 \\
 {\bf SNR NEI}
  & {\bf 0.35} & {\bf 0.26} & {\bf 0.37} & {\bf 0.28} & {\bf 0.021} \\
  
\enddata
\tablecomments{
  All fluxes are in units of 10$^{-12}$ erg cm$^{-2}$s$^{-1}$. $Low S2$ indicates that the Low S2 background fit was used.  
}
\end{deluxetable*}

\begin{deluxetable*}{lllllll}
\tabletypesize{\scriptsize}%
\tablecolumns{7}
\tablewidth{0pc}
\tablecaption{Luminosities and Surface Brightnesses for DEM L50
        \label{tab:lumin50}}
\tablehead{
\colhead{Model}    & \colhead{0.3-2.0 keV}
  & \colhead{0.5-2.0 keV}  & \colhead{0.3-8.0 keV}
  & \colhead{0.5-8.0 keV}  & \colhead{2.0-8.0 keV} & \colhead{Surface Brightness}
  \\
\colhead{(1)} & \colhead{(2)} & \colhead{(3)} 
  & \colhead{(4)} & \colhead{(5)}
  & \colhead{(6)} & \colhead{(7)}
}
\startdata
Center 1T
 & 15.69 & 8.43 & 15.79 & 8.53 & 0.098 & 2.70\\
Center 1T solar
 & 9.41 & 4.93 & 9.51 & 5.03 & 0.099 & $1.63$\\
 Center 2T
  & 21.02 & 10.51 & 21.12 & 10.61 & 0.098 & $3.61$\\
Center 2T solar
 & 12.41 & 5.47 & 12.51 & 5.57 & 0.098 & $2.14$\\
{\bf Center NEI}
& {\bf 10.25} & {\bf 4.95} & {\bf 10.35} & {\bf 5.05} & {\bf 0.099} & {\bf 1.77}\\
Limb 1T
& 14.42 & 6.67 & 14.60 & 6.85 & 0.18 & $1.40$\\
Limb 2T
& 23.81 & 8.25 & 23.98 & 8.42 & 0.18 & 2.31\\
Limb 2T solar
& 10.95 & 3.99 & 11.13 & 4.16 & 0.18 & $1.07$\\
Limb 1T $(Low S2)$
& 13.72 & 9.60 & 13.92 & 9.80 & 0.20 & $1.32$\\
Limb 2T $(Low S2)$
& 12.54 & 8.73 & 12.74 & 8.93 & 0.20 & $1.22$\\
{\bf Limb NEI}
& {\bf 12.54} & {\bf 5.35} & {\bf 12.72} & {\bf 5.53} & {\bf 0.18} & {\bf 1.22}\\
Limb NEI $(Low S2)$
& 12.15 & 8.48 & 12.35 & 8.68 & 0.20 & $1.19$\\
Bright Limb 1T
& 7.72 & 4.50 & 7.77 & 4.54 & 0.047 & 2.81\\
Bright Limb 2T
& 8.50 & 4.33 & 8.54 & 4.38 & 0.047 & $3.08$\\
{\bf Bright Limb NEI}
& {\bf 5.83} & {\bf 3.43} & {\bf 5.87} & {\bf 3.48} & {\bf 0.048} & {\bf 2.12}\\
 SNR 2T 
 & 3.45 & 1.35 & 3.51 & 1.41 & 0.063 & 1.02\\
 SNR 2T solar
& 8.69 & 2.73 & 8.75 & 2.79 & 0.060 & $2.54$\\
 SNR 2T $(Low S2)$
& 1.70 & 1.11 & 1.77 & 1.18 & 0.070 & $0.51$\\
 SNR 2T solar $(Low S2)$
& 10.83 & 3.81 & 10.90 & 3.88 & 0.067 & $3.16$\\
{\bf SNR NEI}
& {\bf 3.67} & {\bf 1.47} & {\bf 3.73} & {\bf 1.54} & {\bf 0.064} & {\bf 1.08}\\
Totals of 1T Models
& 30.11 & 15.11 & 30.39 & 15.38 & 0.27 &$1.87$\\
Totals of 2T Models
& 44.83 & 18.76 & 45.10 & 19.03 & 0.28 &$2.78 $\\
Totals of NEI models
& 22.79 & 10.30 & 23.07 & 10.58 & 0.28 & 1.42 \\
 
\enddata
\tablecomments{
  Luminosities are in units of 10$^{35}$ \ergps. 
  Surface brightness is shown for the $0.3-8.0$ keV band in units of $10^{31}$ erg s$^{-1}$arcsec$^{-2}$. $Low S2$ indicates that the Low S2 background fit was used. 
  Totals are sums of Center and Limb regions, with High S2 background fits, free \nH, and free abundances.
}
\end{deluxetable*}

\begin{deluxetable*}{llllll}
\tabletypesize{\scriptsize}%
\tablecolumns{6}
\tablewidth{0pc}
\tablecaption{Observed Fluxes for DEM L152
        \label{tab:fluxobs152}}
\tablehead{
\colhead{Region and Model}    & \colhead{0.3-2.0 keV}
  & \colhead{0.5-2.0 keV}  & \colhead{0.3-8.0 keV}
  & \colhead{0.5-8.0 keV}  & \colhead{2.0-8.0 keV}
  \\
\colhead{(1)} & \colhead{(2)} & \colhead{(3)} 
  & \colhead{(4)} & \colhead{(5)}
  & \colhead{(6)} 
}
\startdata
{\bf Bubble 1T} 
  &  {\bf 0.93} & {\bf 0.87} & {\bf 0.95} & {\bf 0.89}  & {\bf  0.016} \\
Bubble 2T 
  &   0.93 & 0.88 & 0.94 & 0.89  &  0.017  \\
{\bf Bubble Limb 1T}
 & {\bf 0.77} & {\bf 0.72} & {\bf 0.78} & {\bf 0.73} & {\bf 0.0096} \\
Bubble Limb 2T \nH~fixed
 & 0.77 & 0.73 & 0.78 & 0.74 & 0.011 \\
West 1T  
  &  0.73 & 0.67 & 0.73 & 0.67  &  0.0061 \\
 {\bf West 2T}
  & {\bf 0.73} & {\bf 0.68} & {\bf 0.75} &{\bf  0.70} & {\bf 0.023} \\
West Limb 1T
  & 0.26 & 0.23 & 0.26 & 0.23 & 0.0020 \\
 West Limb 2T \nH~fixed
  & 0.25 & 0.24& 0.26 & 0.24 & 0.0061 \\
 {\bf West Blowout 1T \nH~fixed}
  & {\bf 0.19} & {\bf 0.18} & {\bf 0.20} & {\bf 0.18} & {\bf 0.00070} \\
 West Blowout 2T \nH~fixed
  & 0.20 & 0.18 & 0.20 & 0.18 & 0.00077 \\
 South Bar 1T
  & 0.53 & 0.48 & 0.54 & 0.48 & 0.0053 \\
 South Bar 2T
  & 0.54 & 0.48 & 0.56 & 0.49 & 0.016 \\
 {\bf SNR 1T}
  & {\bf 0.15} & {\bf 0.14} & {\bf 0.16} & {\bf 0.15} & {\bf 0.0042} \\
 SNR 2T
  & 0.15 & 0.15 & 0.16 & 0.15 & 0.0076 \\
 Knot 1 1T solar
  & 0.049 & 0.046 & 0.049 & 0.046 & 0.00019 \\
 Knot 1 1T free abun
  & 0.051 & 0.048 & 0.052 & 0.048 & 0.00036 \\
 Knot 2 1T
  & 0.024 & 0.022 & 0.024 & 0.022 & 0.00012 \\

\enddata
\tablecomments{
  Fluxes are in units of 10$^{-12}$ erg cm$^{-2}$s$^{-1}$. 
}
\end{deluxetable*}

\begin{deluxetable*}{lllllll}
\tabletypesize{\scriptsize}%
\tablecolumns{7}
\tablewidth{0pc}
\tablecaption{Luminosities and Surface Brightnesses for DEM L152
        \label{tab:lumin152}}
\tablehead{
\colhead{Region and Model}    & \colhead{0.3-2.0 keV}
  & \colhead{0.5-2.0 keV}  & \colhead{0.3-8.0 keV}
  & \colhead{0.5-8.0 keV}  & \colhead{2.0-8.0 keV} &\colhead{Surface Brightness}
  \\
\colhead{(1)} & \colhead{(2)} & \colhead{(3)} 
  & \colhead{(4)} & \colhead{(5)}
  & \colhead{(6)} & \colhead{(7)}
}
\startdata
{\bf Bubble 1T} 
  &  {\bf 20.20} & {\bf 9.10} & {\bf 20.26} & {\bf 9.16}  & {\bf  0.054} & {\bf 4.23}\\
Bubble 2T 
  &   22.22 & 10.58 & 22.27 & 10.64  &  0.056  &$4.65$\\
{\bf Bubble Limb 1T}
 & {\bf 17.88} & {\bf 8.61} & {\bf 17.91} & {\bf 8.64} & {\bf 0.032} & {\bf 4.31}\\
Bubble Limb 2T \nH~fixed
 & 17.66 & 8.54 & 17.69 & 8.58 & 0.035 & $4.26$ \\
West 1T  
  &  21.59 & 8.50 & 21.61 & 8.52  &  0.020 & $3.44$\\
 {\bf West 2T}
 & {\bf 22.19} & {\bf 9.12} & {\bf 22.27} & {\bf 9.20} & {\bf 0.075} & {\bf 3.55}\\
 West Limb 1T
  & 6.23 & 2.42 & 6.24 & 2.43 & 0.0066 & $3.26$\\
 West Limb 2T \nH~fixed
  & 7.31 & 3.24 & 7.33 & 3.26 & 0.020 &$3.83$ \\
 {\bf West Blowout 1T \nH~fixed}
  & {\bf 7.22} & {\bf 2.70} & {\bf 7.22} & {\bf 2.71} & {\bf 0.0023} & {\bf 3.07}\\
 West Blowout 2T \nH~fixed
  & 7.26 & 2.79 & 7.27 & 2.79 & 0.0025 &$3.09$\\
 South Bar 1T
  & 12.81 & 4.72 & 12.82 & 4.73  & 0.017 & $3.04$\\
 South Bar 2T
  & 11.99 & 4.29 & 12.04 & 4.34  & 0.050 & $2.85$\\
 {\bf SNR 1T}
  & {\bf 5.72} & {\bf 2.12} & {\bf 5.73} & {\bf 2.14}  & {\bf 0.014} & {\bf 3.22}\\
 SNR 2T
  & 5.72 & 2.16 & 5.75 & 2.18 & 0.025 &$3.23$\\
 Knot 1 1T solar
  & 0.81 & 0.45 & 0.81 & 0.45 & 0.0006 & $3.70$ \\
 Knot 1 1T free abun
  & 0.82 & 0.46 & 0.82 & 0.46 & 0.0012 & $3.74$\\
 Knot 2 1T
  & 0.43 & 0.21 & 0.43 & 0.21 & 0.0004 & $4.02$\\
 Total of 1T Models
  & 54.60 & 22.32 & 54.69 & 22.41  & 0.091 & $3.58$\\
 Total of 2T Models
 & 56.40 & 24.03 & 56.58 & 24.21 & 0.18 & $3.70$\\
 
\enddata
\tablecomments{
   Luminosities are in units of 10$^{35}$ \ergps.
   Surface brightness is shown for the $0.3-8.0$ keV band in units of $10^{31}$ erg s$^{-1}$arcsec$^{-2}$.
   Totals are sums of Bubble, West, and South Bar regions. 
}
\end{deluxetable*}

\begin{deluxetable*}{llllll}
\tabletypesize{\scriptsize}%
\tablecolumns{5}
\tablewidth{0pc}
\tablecaption{Hardness Ratios and Median Energies
        \label{tab:hardness}}
\tablehead{
\colhead{Region}    & \colhead{$Q$ (1T fits)} & \colhead{$Q$ (2T fits)} &\colhead{$Q$ (NEI fits)}
 & \colhead{$Q$ (count rate)} & \colhead{Median Energy (keV)}
    \\
    \colhead{(1)} & \colhead{(2)} & \colhead{(3)} & \colhead{(4)} & \colhead{(5)} & \colhead{(6)}
}
\startdata
 DEM L50 Center
  & -0.67\tablenotemark{a} & -0.72\tablenotemark{a} & -0.63\tablenotemark{a} & ... & 0.723\\
 DEM L50 Limb
  & -0.60\tablenotemark{a} & -0.73\tablenotemark{a} & -0.63\tablenotemark{a} &... & 0.723\\
 DEM L50 Bright Limb
  & -0.48\tablenotemark{a} & -0.55\tablenotemark{a} & -0.43\tablenotemark{a} & ... & 0.767\\
 DEM L50 Center + Limb
  & -0.63\tablenotemark{a} & -0.72\tablenotemark{a} & -0.63\tablenotemark{a} & 0.46 $\pm$ 0.02 & 0.723\\
 DEM L50 SNR
  & ... & -0.56\tablenotemark{a} & -0.54\tablenotemark{a} & 0.32 $\pm$ 0.02 & 0.694\\
 DEM L152 Bubble
  & -0.34 & -0.35 & ... & 0.75 $\pm$ 0.02 & 0.869\\
 DEM L152 Bubble Limb
  & -0.31 & -0.33 & ... & ... & 0.854\\
 DEM L152 West
  & -0.50 & -0.51 & ... & 0.67 $\pm$ 0.02 & 0.825 \\
 DEM L152 West Limb
  & -0.48 & -0.49 & ... & ... & 0.810\\
 DEM L152 West Blowout
  & -0.58 & -0.59 & ... & ... & 0.759\\
 DEM L152 South Bar
  & -0.50 & -0.53 & ... & 0.63 $\pm$ 0.03 & 0.825\\
 DEM L152 SNR
  & -0.51 & -0.50 & ... & 0.80 $\pm$ 0.07 & 0.942\\
 
\enddata
\tablecomments{
   The hard energy band is $E=0.6 - 1.6$ keV, and the soft band is
$E=0.3 - 0.6$ keV. 
}
\tablenotetext{a}{Using free abundance fits with the High S2 background.}

\end{deluxetable*}
\subsection{DEM L50}
\label{sec:deml50models}

\begin{figure*}[!ht]
\epsscale{1.0}
\plotone{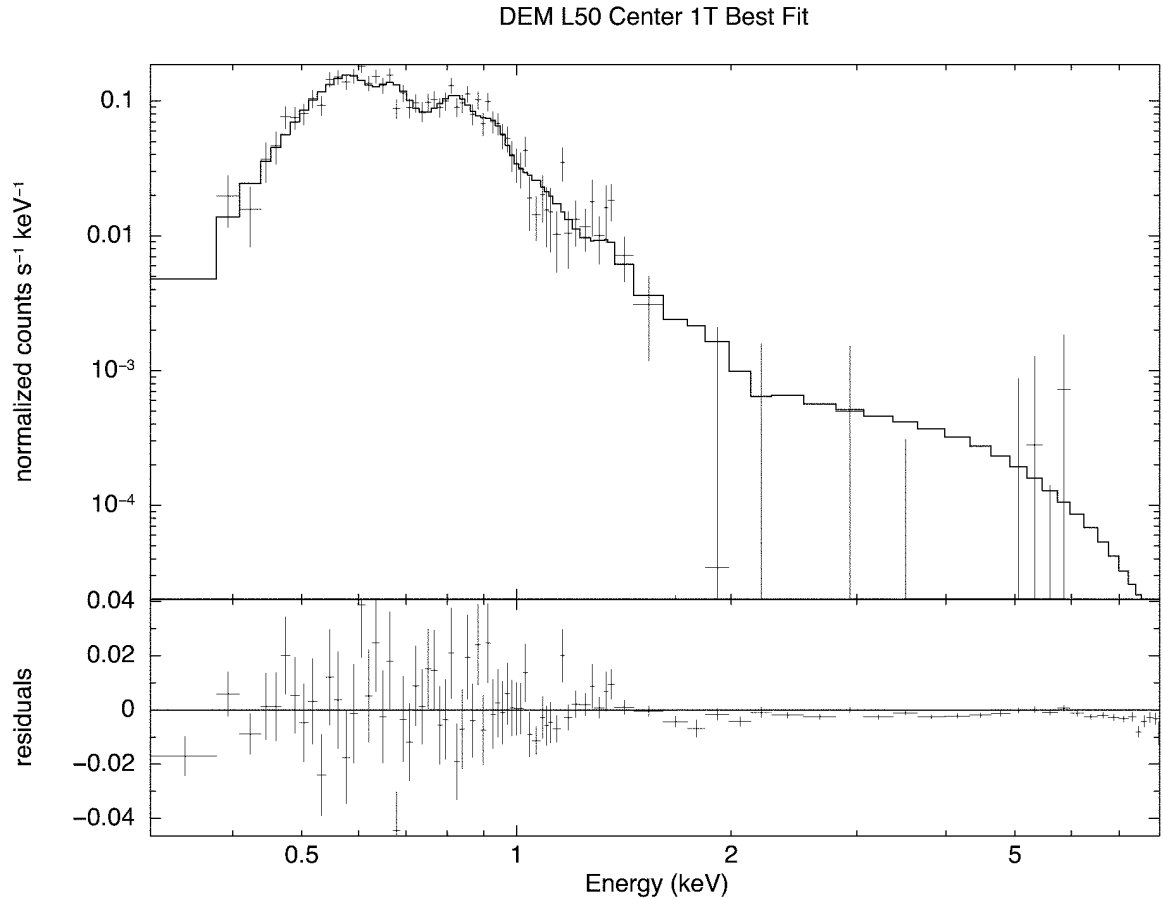}
  \caption{The X-ray spectrum of the Center region of DEM L50, with the one-temperature thermal plasma best fit and residuals (see Table~\ref{tab:fit50}). Data have been re-binned for clarity. }
 \label{fig:50spectra}
\end{figure*}

Modeled fits to DEM L50 are shown in Tables~\ref{tab:fit50} and~\ref{tab:fit50ne}. The fits in Table~\ref{tab:fit50} model the emission as one- and two-temperature thermal plasmas. The best fit \nH~in Column 3 is consistent with the observations of the \hi~in DEM L50 by \citet{oey02}. The first temperature component, $kT_1$, appears in Column 4, with its normalization in Column 5, and the second component, $kT_2$, appears in Column 6, with its normalization in Column 7. Column 8 shows the ``Effective kT", defined as the average $kT$ of the region weighted by the normalization. The ``Effective kT" thus shows the dominant temperature component in the region. Columns 9 and 10 give the abundances relative to solar, with the $\alpha$/Fe ratio in Column 11. Columns 12 and 13 show the reduced $\chi^2$ and the number of degrees of freedom of the fit. The statistically best fits, determined by running {\sc ftest} in XSPEC, are shown in bold. Figure~\ref{fig:50spectra} shows a sample fit to the Center region of DEM L50.

From the best fits for the Limb and Center regions in Table~\ref{tab:fit50}, the Center's temperature distribution appears more homogeneous than the Limb's, and the two-temperature best-fit to the Limb shows a higher temperature component that may correspond to the Bright Limb subregion. This correlation would be consistent with shock-heating of the southern part of the Limb by an expanding SNR, while the rest of the Limb remains at a cooler temperature.

\subsubsection{Low $\alpha$/Fe vs. non-equilibrium ionization}
\label{sec:deml50_abun}
In each region of DEM L50, spectral fits consistently gave low $\alpha$/Fe ratios (see Table~\ref{tab:fit50}), with typical values of about 0.3 compared to an average LMC value of 0.6 \citep{smith99}. These low ratios are especially surprising, given that we might have expected a high $\alpha$/Fe ratio due to an enhancement in $\alpha$ elements from core-collapse SNe. The bubble has likely experienced $\sim 2$~SNe \citep{oey96c}, so an $\alpha$ enhancement would be natural. Raising the $\alpha$/Fe ratio by forcing the abundances to solar values resulted in the models labeled `solar' in Table~\ref{tab:fit50}. In every case, {\sc ftest} showed that the free abundance fits were statistically better than the solar values. The {\sc ftest} command calculates a {\it p}-value, which here denotes the probability that solar-abundance material could randomly produce a spectrum fit by the more complicated, free abundance fit. The resulting {\it p}-values ranged from 0.049 to as low as $4\times10^{-6}$, showing that the solar fits have a very low probability of being correct. Although the solar abundances are statistically excluded, there may be systematic problems in forward-fitting low-resolution X-ray spectra, which could give us misleading estimates of the abundances and their uncertainties. Using the Mekal or Raymond-Smith models instead of APEC did not affect the $\alpha$/Fe ratios. 

\begin{figure*}[!ht]
\epsscale{1.0}
\plotone{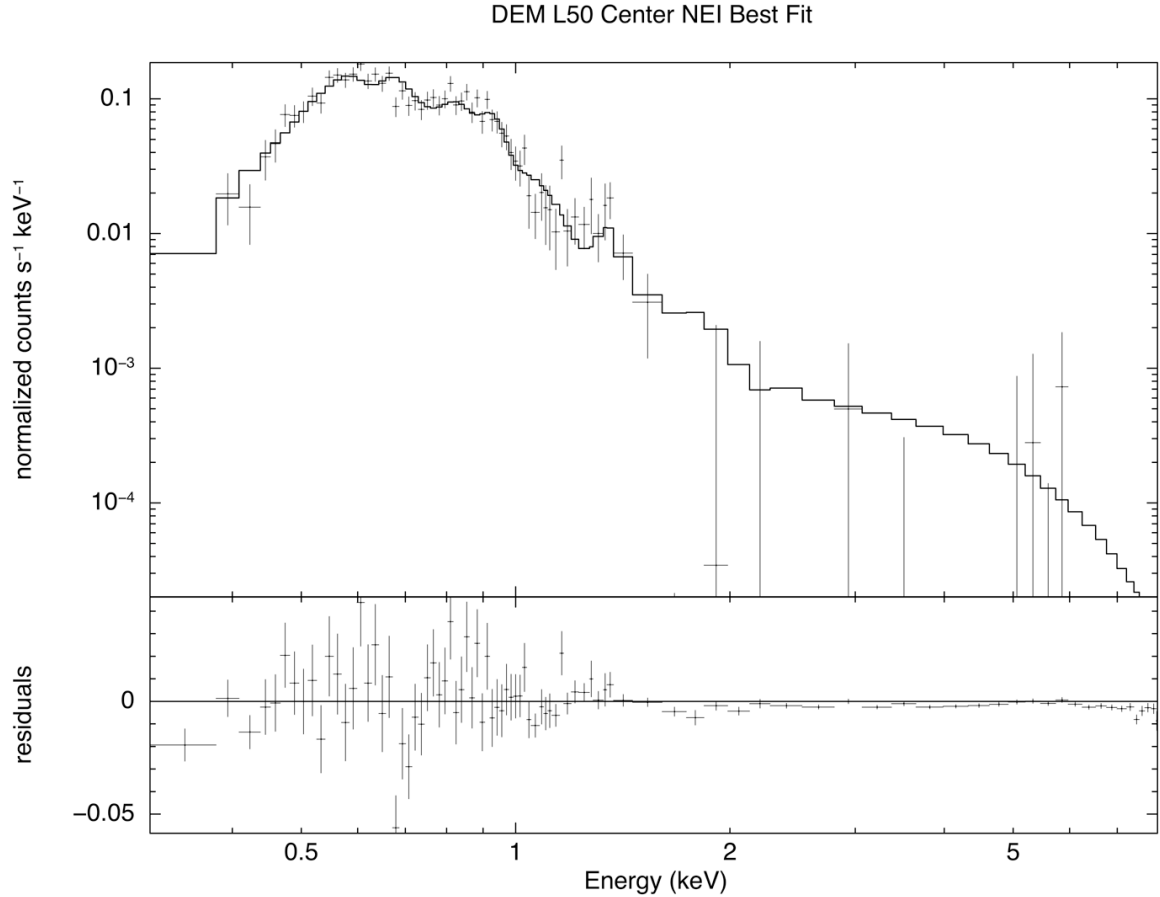}
 \caption{The X-ray spectrum of the DEM L50 Center region, with the non-equilibrium ionization fit and residuals shown (see Table~\ref{tab:fit50ne}). Data have been re-binned for clarity.}
  \label{fig:50spectra_nei}
 \end{figure*}

Another possibility is that collisional ionization equilibrium is not a valid assumption for DEM L50.  To test this scenario, we performed the non-equilibrium ionization fits shown in Table~\ref{tab:fit50ne}. Temperature is shown in Column 4, the normalization is in Column 5, abundances relative to solar are in Columns 8 and 9, and the reduced $\chi^2$ and number of degrees of freedom are in Columns 11 and 12, respectively. The $\alpha$/Fe ratios found with these new fits are all consistent with solar values. A sample non-equilibrium ionization fit is shown in Figure~\ref{fig:50spectra_nei} for the DEM L50 Center region.

To see whether or not non-ionization equilibrium is a plausible scenario, we consider the ionization timescale. The ionization timescale in seconds is found by dividing $\tau$, the density-weighted ionization timescale in s cm$^{-3}$ (Column 6), by the electron density, $n_{e}$. To find the electron density, we used 
\begin{equation}
\label{eqn:dens}
\eta = \frac{10^{-14}}{4\pi D_{A}^2}\int n_{e} n_{\rm H} dV
\end{equation}
where $\eta$ is the normalization of the fit, in units of cm$^{-5}$ and $D_A$ is the distance to the superbubble, assumed to be 50 kpc. We also assumed a spherical geometry, uniform density distribution, and $\sim$10\% He abundance ($n_{e}\sim1.2~n_{\rm H}$; see, e.g., \citealt{peimbert07}). The resulting ionization timescale estimates are shown in Column 7 of Table~\ref{tab:fit50ne} and are all on the order of $10^5$ years, about 10\% of the superbubble's age. While our observation of the bubble in this stage would be a chance occurrence to some extent, this timescale is not unrealistically brief and a recent supernova is also consistent with many of the observed properties of the bubble as we will discuss later. Furthermore, the hypothesized brightening from SNR impacts should enhance the detectability of bubbles with recent SNe, increasing the likelihood of observing a bubble that is not in equilibrium.

If the low $\alpha$/Fe abundances are indeed real, they may reflect an Fe enrichment in the ISM near DEM L50. DEM L50 is located in a large, 1.4 kpc diameter void in the ISM \citep{oey02}; a single Type Ia supernova could not noticeably enrich this entire region, nor could it enrich the estimated $1.7\times10^5$ \Msol~of material swept up by DEM L50 \citep{oey02}. On the other hand, if the density is low enough, the interior of DEM L50 could be enriched by a single Type Ia event, and SNR N186 D near the bubble would be a likely candidate. The $\alpha$/Fe ratio of the SNR, however, does not seem noticeably different from the rest of the bubble (see \S~\ref{sec:SNRmodels}) and may in fact be less Fe-enriched than the Center region. There is no clear spectral contrast between the SNR and regions dominated by the shocked ISM like that seen in the Type Ia SNR DEM L71 \citep{hughes03}, nor are there any  noticeable Fe, S, or Si features such as those seen in Type Ia SNR N103B \citep{hughes95}. Furthermore, SNR N186D's proximity to young stars and star-forming regions make a core-collapse SN origin more likely \citep[\eg][]{chu88}.

Dust destruction from a supernova shock wave cannot explain the low $\alpha$/Fe ratio. Although dust destruction would increase the Fe abundance, the destruction of silicates would concomitantly raise the O abundance (Compi{\`e}gne, M., private communication; see \citeauthor{whittet03} 2003 for O and Fe abundances in dust). We note that DEM L50 does have an uncommon infrared morphology (Slater et al. 2010, in prep.); its 8 $\mu$m emission appears to fill the bubble's H$\alpha$ shell, whereas many other LMC superbubbles only show polycyclic aromatic hydrocarbon (PAH) emission external to the bubble. This morphology is likely explained by PAH emission on the interface between the ionized and the foreground neutral layer, which is seen in projection toward the center of the bubble (see Fig.~\ref{fig:img_deml50} and \S~\ref{sec:images:deml50}). An unusual dust environment does not appear to be the cause of the low observed $\alpha$/Fe ratio.

Similarly odd Fe abundances have been observed in other objects. DEM L316 consists of a pair of interacting SNRs; one SNR has a high Fe abundance while the other does not \citep{williamschu05,nishiuchi01}. As mentioned previously, the LMC $\alpha$/Fe ratio is low in general, $\sim$0.6 \citep{smith99}; DEM L50 may simply lie in a region of the ISM with a lower than average $\alpha$/Fe ratio.

Of course, the $\alpha$/Fe ratio we observe could also result from inadequate X-ray models or spectra, rather than genuine properties of the emitting material. In X-ray spectral models, parameters such as temperature and column density are degenerate with the metallicity, making the model difficult to constrain \citep{dahlem00}. A variety of X-ray models may fit the same spectrum equally well. For instance, other authors have found that the need for unusual abundances in a spectral fit may disappear when a more complicated, multi-component X-ray spectral model is used instead \citep[\eg][]{weaver00, strickland02}. The $\sim$0.12 keV resolution of {\it Chandra} ACIS spectra and model degeneracies make it unclear which precise combination of models should be used, however. Determining the origin of the unusual abundances in DEM L50 requires further investigation of both the LMC ISM and the X-ray models used, and the explanation of the observed abundances may have to wait for higher resolution spectra. Our current data suggest, however, that the non-equilibrium ionization scenario is the most plausible explanation for DEM L50's low observed $\alpha$/Fe ratio. 
\subsubsection{Luminosities}
\label{sec:deml50_lumin}
Luminosities for all regions and models are shown in Table~\ref{tab:lumin50}, with the non-equilibrium ionization fits shown in bold. The Limb contributes roughly 50\% of the total emission. The Bright Limb subregion provides 13-41\% of the total. Hardness ratios and median energies appear in Table~\ref{tab:hardness} and indicate that the Bright Limb is clearly spectrally harder than the Limb as a whole, while the Limb and Center have comparable hardness ratios. This relative spectral hardness suggests that the Bright Limb was recently heated. For the two-temperature equilibrium fit to the Limb, the Limb appears to have a softer hardness ratio than the Center. The two-temperature fit has a slightly higher absorbing column density and an additional lower temperature component, which account for the difference. The temperatures of the non-equilibrium fits are less precise, and the larger error bars make the temperature differences between the regions unclear. The relative hardness of the Bright Limb is still apparent when the non-equilibrium fits are considered, however, and is confirmed by the median energies of the regions. 

As was noted in  \S~\ref{sec:images:deml50}, absorption due to intervening \hi~may be responsible for the apparent limb-brightening observed in DEM L50. When the areas of the extracted regions are considered, the Center does appear to have a $\sim$20-45\% higher surface brightness than the Limb as a whole (Table~\ref{tab:lumin50}). The Bright Limb subregion, however, has a surface brightness comparable to or greater than the Center. The \hi~column density has a significant effect on the calculated luminosity, however, and as can be seen in Table~\ref{tab:lumin50}, the luminosity varies substantially depending on the model. The \nH~error bars in our models correspond to typical uncertainties in the X-ray luminosity of about 25\%, but the uncertainty is up to 50\% for the Bright Limb. When Galactic \nH~is included, the \nH~contours observed by \citet{oey02}, shown in Figure~\ref{fig:img_deml50}, do agree with the modeled \nH, given the error bars in the fits, \nH~variations within the regions, and uncertainties in the \nH~observations. Nevertheless, we cannot conclusively determine whether the apparent brightness of the Bright Limb is genuine or due to line-of-sight absorption.

The temperature and hardness of the Bright Limb subregion support the idea that an off-center SNR heated the shell wall in that region, while the Center's brightness is most likely a result of mass-loading due to conductive evaporation from the shell walls. Mass-loading from swept-up clouds is unlikely given DEM L50's location in a void in the ISM and the lack of observed clouds in X-ray or other wavelengths (see Fig.~\ref{fig:rgb}). We discuss this further in \S~\ref{sec:discussion:massloading}. Metallicity enhancement is also not observed; while Fe is enhanced, the $\alpha$ elements, which should have the greatest effect on the cooling function, are not. Given the contribution of the Limb to the total luminosity and the lack of observed clumps in the ISM near DEM L50 (see Fig.~\ref{fig:rgb}), an off-center SNR and thermal conduction from the shell walls into the center are a more viable explanation for DEM L50's brightness than ablation or evaporation of ISM cloudlets.

\subsection{DEM L152}
\label{sec:deml152models}

\begin{figure*}[!ht]
\epsscale{1.0}
\plotone{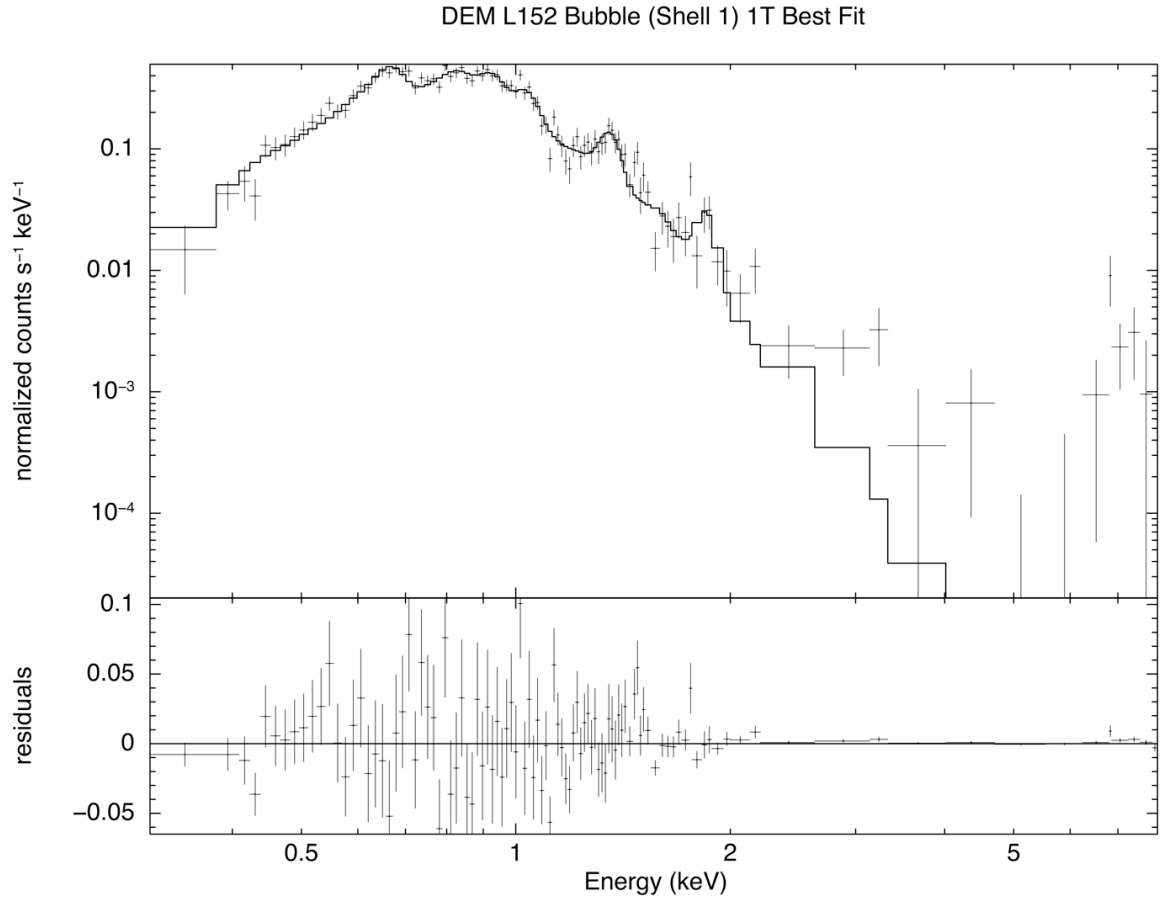}
 \caption{The X-ray spectrum of the Bubble (Shell 1) region of DEM L152, with the one-temperature thermal plasma best fit and residuals (see Table~\ref{tab:fit152}). Data have been re-binned for clarity.}
 \label{fig:152spectra}
 \end{figure*}

Like DEM L50, DEM L152 is an X-ray-bright superbubble, although as seen in Figures~\ref{fig:img_deml50} and~\ref{fig:img_deml152}, its \hi~environment and X-ray and optical morphology differ substantially from that of DEM L50. Consequently, DEM L152's derived X-ray properties also generally differ from DEM L50; the characteristics the two objects have in common may help resolve the question of why some superbubbles become X-ray-bright. We modeled the emission from DEM L152 with one- and two-temperature thermal plasmas. Table~\ref{tab:fit152} reports the resulting fits; the column headings are the same as in Table~\ref{tab:fit50}, except that the background fit, which is the same for all fits, is not listed. In the case of the West Limb and South Bar, it is not clear which fit is statistically better, so neither fit is shown in bold. Figure~\ref{fig:152spectra} shows the best fit to the Bubble region as an example.

DEM L152 is a few $10^6$ K hotter than DEM L50 on average, a result of the higher number of O stars and supernovae in DEM L152 \citep{oey95,oey96b}. The Bubble Limb appears slightly cooler than the Bubble region as a whole, which is reasonable since it makes up the outer edge of the Bubble and likely consists of denser, cooler material. The West region shows a 0.25 keV component as well as an extremely hot 0.79 keV component. This hot component appears in the West Limb region two-temperature fit, but not the West Blowout, and may also account for a substantial portion ($\sim$25\%) of the emission from the South Bar. The localization of this hot component to the adjacent West Limb and South Bar regions suggests recent heating in this area, perhaps from a supernova explosion. An expanding SNR would also explain the limb brightening seen in the West region. Interestingly, \citet{chu90} also suggest the presence of an off-center supernova remnant in the southwest due to the X-ray brightness in the South Bar, and \citet{magnier96} find high-velocity gas in the region.

As expected, the two blowout regions are slightly cooler than the Bubble region. Most of the South Bar's emission is fit by a 0.29 keV component, while the hot component mentioned above accounts for the remainder. Likewise, a 0.25 keV plasma characterizes the West Blowout.  The blowouts are cooler than the 0.37 keV Bubble region and cooler than (or at least comparable to) the 0.33 keV Bubble Limb region, presumably having cooled due to adiabatic expansion. 

Abundances for each region show enrichment from core-collapse supernovae (see Table~\ref{tab:fit152}), which raises the $\alpha$/Fe ratio. We discuss the implications of this enrichment for the expected luminosity in~\S~\ref{sec:discussion:metallicity}. 

\begin{figure}[!ht] 
\epsscale{1.0}
\plotone{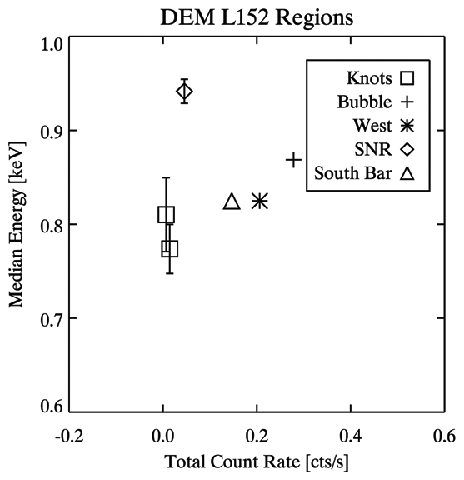}
  \caption{Median Energy vs. Total Count Rate for regions in DEM L152. The plot shows rough spectral differences between the regions. Error bars were estimated using the quantile error estimation approach outlined in Appendix B of \citet{hong04}.
  }
  \label{fig:knotplot}
\end{figure}

Two clumps of X-ray emitting material, a few parsecs in diameter, appear in the South Bar region (see Fig.~\ref{fig:ext_regions}). We extracted and fit spectra for these two knots (see Table~\ref{tab:fit152}), fixing the column density to the best fit value for the South Bar region. Due to low signal-to-noise, we only used single temperature fits and fixed abundances to solar values for the second knot. The resulting luminosities should be viewed as estimates only. Nevertheless, the luminosities are one to two orders of magnitude lower than the luminosity of the other regions and clearly do not have much effect on the total emission.  We plot the median energy versus total count rate for the knots and regions of DEM L152 in Fig.~\ref{fig:knotplot}. The knots do not exhibit any detectable spectral difference relative to the rest of the superbubble. 

The luminosities for each region and subregion of DEM L152 are shown in Table~\ref{tab:lumin152}; the \nH~error bars correspond to an uncertainty of $\sim$15\% in the calculated luminosities. The Bubble region provides $\sim$~37\% of the total X-ray luminosity (33\% from the Bubble Limb, 4\% from diffuse emission), the West region contributes $\sim$~41\% (14\% from the West Limb, 13\% from the West Blowout, and 14\% from diffuse emission), and the South Bar provides $\sim$~21\% (19\% from diffuse emission, 2\% from the knots). The intrinsic surface brightnesses are unclear, as they depend on the model and column density assumed. For example, the best fits for the Bubble region imply some slight limb-brightening, while the two-temperature fits indicate the opposite. If the limb-brightening observed in Fig.~\ref{fig:rgb} is not intrinsic, it must instead result from higher absorption towards the Bubble center. This hypothetical absorption would conflict with the \hi~contours shown in Fig.~\ref{fig:img_deml152}, however, which show a lower column density towards the center of the Bubble. Thus, we expect that the observed limb-brightening in the Bubble region is real. 

A comparison of the hardness ratios and median energies in Table~\ref{tab:hardness} shows that the South Bar and West Blowout regions are softer than the rest of the superbubble, consistent with a blowout origin and with the previous X-ray observations of \citet{chu93} and \citet{magnier96}. This variation in hardness can also be seen in Figure~\ref{fig:rgb}e. The Bubble and Bubble Limb regions are noticeably harder than the other regions of DEM L152. In general, we find softer hardness ratios for DEM L152 than previous authors \citep[\eg][]{wang91}, a result of fitting the column density instead of assuming a fixed conversion between count rate and luminosity. 

As with DEM L50, limb regions account for roughly half of the observed X-ray emission. The South Bar and West Limb regions alone provide a third of the total emission and may be associated with a recent off-center SNR as discussed above. In contrast, the two knots make up only 2\% of the total X-ray emission, suggesting that emission from clumps cannot explain DEM L152's X-ray luminosity.

\subsection{SNRs}
\label{sec:SNRmodels}

The SNR N186 D appears at the northern edge of DEM L50. The fact that N186 D is detected in \oiii~while DEM L50 is not suggests they may be two physically distinct objects \citep{lasker77}; similar \hi~kinematics in both objects, however, may indicate that the SNR is associated with the larger bubble \citep{oey02}, and thus the exact relationship between the SNR and the superbubble remains unclear. 

SNR N186 D's X-ray spectra can be fit by a non-equilibrium model or a two-temperature equilibrium model. With free abundances, the equilibrium fits show a 0.19 keV component and a weaker 0.72 keV component. These temperatures are typical of SNRs in the LMC \citep[see][]{williams99} and in our own Galaxy \citep[\eg][]{temim09,hui09}. If abundances are forced to solar values, however, the two temperature components are 0.08 keV and 0.29 keV. The dominant 0.08 keV component is a lower temperature than any of the LMC remnants studied by \citet{williams99} and even the 0.29 keV component is on the lower end of her sample. The free abundance fit is statistically better than this solar fit, with a {\it p}-value of 0.05, and shows the same low $\alpha$/Fe ratio seen in the rest of DEM L50. Although we cannot rule out the possibility that this SNR is a Type Ia remnant, the consistency of the $\alpha$/Fe ratio throughout DEM L50 suggests the abundances are associated with the ISM environment of DEM L50 or alternatively, that the SNR is not in ionization equilibrium. As discussed in ~\S~\ref{sec:deml50_abun}, the non-equilibrium scenario is the more likely explanation. A non-equilibrium ionization fit gives an ionizing timescale of 11,000 years, a temperature of 0.58 keV, and an $\alpha$/Fe ratio of 0.64, close to the typical LMC value of $\sim$0.6 \citep{smith99}. 

The non-equilibrium ionization model gives a total X-ray luminosity in the $E=0.3-8.0$ keV band of $3.73\times10^{35}$ \ergps. The luminosity varies by an order of magnitude among the potential models, however, from $2\times10^{35}$ to $1\times10^{36}$ \ergps, a result of the higher \nH~in the models with solar abundances. These possible luminosities are all within the range of $\sim10^{34}$-$10^{37}$ \ergps found by \citet{williams99b} for SNRs in the LMC and by \citet{long10} for M33 SNRs. 

The median energy for SNR N186 D, shown in Table~\ref{tab:hardness}, is softer than its hardness ratio would suggest. The \hi~appears patchy in this region of DEM L50, and the observations by \citet{oey02} show a substantial gradient in the \hi~column density near the SNR. These \nH~variations may be affecting our modeled temperatures in this region. 

DEM L152's SNR, 0523-679 \citep{chu93}, consists of a 0.38 keV plasma, close to the median temperature value of Williams' sample of LMC SNRs \citep{williams99}. The SNR shows a high $\alpha$/Fe ratio characteristic of core-collapse SNe. The extremely high and unconstrained value of the SNR's $\alpha$/Fe ratio is a result of the lower signal-to-noise in this small region of the bubble. The total X-ray luminosity is $5.73\times10^{35}$ \ergps, typical of other SNRs \citep[\eg][]{williams99b,long10,seward10}. The SNR does appear to differ spectrally from the rest of the bubble, as seen in Fig.~\ref{fig:knotplot}. 

\section{Model comparisons}
\label{sec:models}

\subsection{Weaver model}
\label{sec:models:weaver}

We modeled the central temperature and X-ray luminosity in the [0.3-2.0 keV] band for DEM L50 and DEM L152 using the similarity solutions of \citet{weaver77} as implemented by \citet{oey95}. Energy and mass input were not assumed to be constant; time-dependent energy input from stellar winds and SNe was calculated using the mass loss rates from \citet{schaerer93} and the stellar population shown in Table~\ref{tab:stelpop} (see \citealt{oey95} for details). 

For DEM L50, we assumed an ambient density of 1.4 \pcc as explained in \citet{oey96c}, a metallicity of 0.4 \Zsol~and an age of 5 Myr, at which point 2 SNe are expected to have occurred if a \citet{salpeter55} initial mass function (IMF) is assumed \citep{oey96c}. We modeled DEM L152 using a density of 2.5 \pcc \citep{oey95}, a metallicity of 0.4 \Zsol~and an age of 6 Myr, incorporating the effects of 4 SNe as suggested in \citeauthor{oey95}'s (1995) analysis of DEM L152's mass function. The resulting X-ray luminosities and central temperatures are shown in Table~\ref{tab:models} and compared with the observed values. The observed X-ray luminosities are an order of magnitude higher than the standard model predicts. Varying the input ambient density or metallicity did increase the X-ray luminosity, but even increasing these parameters by an order of magnitude still resulted in an X-ray luminosity lower than the observed value. 

The radii calculated with the Weaver model likewise do not match the observed values; both bubbles have predicted radii of more than 100 pc (130.6 pc for DEM L50, 146.6 pc for DEM L152), whereas the observed radii are around 50 pc. This ``growth-rate discrepancy" has been observed in bubbles with a range of sizes, from bubbles around individual Wolf-Rayet stars to the superbubbles we consider here (\eg~\citealt{saken92,brown95,oey95,cappa03}; see review by \citealt{oey09}). One possible explanation for the smaller observed radii is that excess radiative cooling has caused the bubble to lose more energy than predicted. To see if the high X-ray luminosities of our objects provide enough cooling to slow the bubble's growth, we use the formula for the bubble radius given by \citet{weaver77}:

\begin{equation}
\label{eqn:radius}
R=27n_0^{-1/5}L_{36}^{1/5}t_6^{3/5} pc ,
\end{equation}
where $n_0$ is the ambient density in cm$^{-3}$, $L_{36}$ is the mechanical luminosity in $10^{36}$ \ergps, and $t_6$ is time in $10^6$ years. Assuming that supernovae dominate the mechanical energy input, we can estimate the mechanical luminosity as the energy provided by SNe divided by the age of the superbubble. Two SNe and an age of 5 Myr (DEM L50) or four SNe and an age of 6 Myr (DEM L152) yield a mechanical luminosity of order $10^{37}$ \ergps. The X-ray radiative losses for both bubbles, on the other hand, are on the order of $10^{36}$ \ergps. Although our superbubbles are X-ray-bright, the observed X-ray luminosity is therefore insufficient to account for the small radius observed in both bubbles. 

\subsection{Hydrodynamic Model}
\label{sec:models:hd}

In contrast to the Weaver model, the one-dimensional hydrodynamic model of DEM L50 from \citet{oey04b} does correctly fit the radius and velocity of the bubble. The mechanical input is the same as before, but this model assumes a higher ambient pressure of $P/k=1\times10^5$ cm$^{-3}$ K, neglects thermal conduction at the shell walls, and includes the effect of a SNR impact on the shell.  While the higher ambient pressure results in the correct radius, the X-ray luminosity predicted by this model is far too low, an order of magnitude lower than the Weaver model, which was itself lower than the observed luminosity. This low luminosity suggests that thermal conduction does play a role in DEM L50 and is not strongly suppressed by magnetic fields. If thermal conduction is occurring, the bubble temperature should be lowest at the outer edge, where dense, cold evaporated material is present, and should increase toward the bubble center. The equilibrium fits to the DEM L50 Limb and Center regions do show this temperature contrast; the effective temperature of the Limb is only 0.09 keV, while the value  increases to 0.15 keV for the Center region. The non-equilibrium fits, on the other hand, suggest that the Limb is slightly hotter than the Center. The possibility of a cooler Limb is within the error bars, however, and the contribution of the heated Bright Limb region may raise the average temperature of the Limb. 

\begin{figure}[!ht] 
\epsscale{1.0}
\plotone{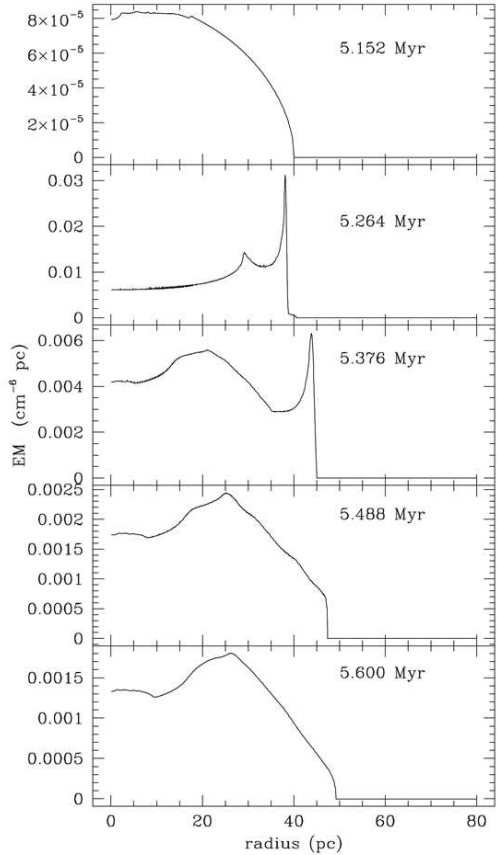}
  \caption{Emission measure along the line of sight for the 1-D hydrodynamic models of DEM L50 by \citet{oey04b}. The top panel shows the modeled bubble at t=5.152 Myr, before the SN at 5.26 Myr goes off. The subsequent panels show the bubble after the SNR expands into the shell walls.
  }
  \label{fig:modelprofile}
\end{figure}

The X-ray luminosity at different times is shown in Table~\ref{tab:hydro} and compared with the Weaver model luminosity for DEM L50 and the observed luminosity. A supernova goes off at $t$=5.26 Myr, and the SNR increases the bubble's luminosity by several orders of magnitude. The emissivity profiles of the bubble before and after the SNR hits the shell walls are shown in Figure~\ref{fig:modelprofile}. The SNR impact enhances the emissivity of the bubble's limb and also increases the shell velocity from 1 \kmps to 23 \kmps in agreement with the observed value. Although the luminosities of this model are unrealistic, the response of the luminosity and shell velocity to the SNR impact suggests that SNR impacts may be important in our superbubbles.    

Neither the Weaver model nor the 1-D hydrodynamic model is able to match the observed luminosities of the superbubbles. The hydrodynamic model suggests, however, that mass-loading via thermal evaporation from the shell walls must play a role in these bubbles. Mass-loading from the shell walls and heating from SNR impacts are consistent with our observations and could account for most of the superbubbles' X-ray emission. 

\begin{deluxetable*}{llllllll}
\tabletypesize{\scriptsize}%
\tablecolumns{8}
\tablewidth{0pc}
\tablecaption{Stellar Populations
        \label{tab:stelpop}}
\tablehead{
\colhead{Bubble}    & \colhead{85 \Msol} & \colhead{60 \Msol}
  & \colhead{40 \Msol}  & \colhead{25 \Msol}
  & \colhead{20 \Msol} 
  & \colhead{15 \Msol} & \colhead{12 \Msol}
  \\
\colhead{(1)} & \colhead{(2)} & \colhead{(3)} 
  & \colhead{(4)} & \colhead{(5)}
  & \colhead{(6)} & \colhead{(7)}
  & \colhead{(8)}
}
\startdata
DEM L50  & 1 & 1 & 3 & 1 & 8 & 6 & 9 \\
DEM L152  & 1 & 3 & 4 & 11 & ... & 12 & ... \\
\enddata
\tablecomments{
  Number of stars of a given mass assumed for the Weaver-based models of DEM L50 and DEM L152. Data from \citet{oey95} and \citet{oey96c}.
}
\end{deluxetable*}

\begin{deluxetable*}{lllllll}
\tabletypesize{\scriptsize}%
\tablecolumns{7}
\tablewidth{0pc}
\tablecaption{Weaver Model Results
        \label{tab:models}}
\tablehead{
\colhead{Bubble Model}    & \colhead{$n_0$} & \colhead{Age}
  & \colhead{$L_{x}$[0.3-2.0 keV]}  & \colhead{$kT_{c}$}
  & \colhead{$L_{x}$[0.3-2.0 keV],obs\tablenotemark{a}} 
  & \colhead{$kT_{c}$,obs}
  \\
  \colhead{(1)} & \colhead{(2)} & \colhead{(3)} 
  & \colhead{(4)} & \colhead{(5)}
  & \colhead{(6)} & \colhead{(7)}
}
\startdata
DEM L50 Weaver & 1.4\pcc & 5 Myr & 1.31 $\times 10^{35}$\ergps & 0.27 keV & $2.0-4.5\times 10^{36}$\ergps & 0.15 keV\tablenotemark{b} or 0.28 keV\tablenotemark{c} \\
DEM L152 Weaver & 2.5\pcc & 6 Myr & 5.05 $\times 10^{35}$\ergps & 0.32 keV & $5.4-5.7\times 10^{36}$ \ergps & 0.37 keV\tablenotemark{d} \\
\enddata
\tablenotetext{a}{Excludes SNRs.}
\tablenotetext{b}{From single-temperature equilibrium fit to Center region.}
\tablenotetext{c}{From non-equilibrium fit to Center region.}
 \tablenotetext{d}{From best fit to Bubble region.}
 
\end{deluxetable*}

\begin{deluxetable*}{llllllll}
\tabletypesize{\scriptsize}%
\tablecolumns{8}
\tablewidth{0pc}
\tablecaption{Hydrodynamic Model Luminosities
        \label{tab:hydro}}
\tablehead{
\colhead{Energy Band} & \colhead{$t=$5.152 Myr}    & \colhead{$t=$5.264 Myr} & \colhead{$t=$5.376 Myr}
  & \colhead{$t=$5.488 Myr}  & \colhead{$t=$5.600 Myr}
  & \colhead{$L_{x}$ Weaver} 
  & \colhead{$L_{x}$ Obs\tablenotemark{a}}
  \\
\colhead{(1)} & \colhead{(2)} & \colhead{(3)} 
  & \colhead{(4)} & \colhead{(5)}
  & \colhead{(6)} & \colhead{(7)} & \colhead{(8)}
}
\startdata
0.3-2.0 keV & $3.7\times10^{31}$ & $1.2\times10^{34}$ & $8.9\times10^{32}$ & $6.0\times10^{33}$ & $3.0\times10^{33}$ & $1.3\times10^{35}$ & $2.0-4.5\times10^{36}$ \\
2.0-8.0 keV & $1.3\times10^{31}$ & $2.5\times10^{33}$ & $1.1\times10^{33}$ & $5.3\times10^{32}$ & $3.9\times10^{32}$ & ... & $2.8\times10^{34}$ \\
\enddata
\tablecomments{Results refer to DEM L50 only.}
\tablenotetext{a}{Excludes SNRs. Luminosities in units of \ergps.}
\end{deluxetable*}

\section{Additional X-ray Enhancement Mechanisms}
\label{sec:discussion}

\subsection{Metallicity}
\label{sec:discussion:metallicity}

 \citet{silich01b} suggests that since emission lines of $\alpha$ elements dominate the soft X-ray emission, metallicity enhancement from SNe could explain the anomalously high X-ray luminosities observed in many superbubbles. As discussed earlier (\S\S~\ref{sec:deml50models}-\ref{sec:deml152models}), DEM L152 shows an $\alpha$ enhancement, but DEM L50 may not, indicating that a different explanation would be required for that bubble.  
 
Assuming the initial stellar population proposed by \citet{oey95}, approximately four SNe have gone off in DEM L152, from one 85\Msol~progenitor and three 60\Msol~progenitors (see Table \ref{tab:stelpop}). These SNe should input a total 65\Msol~of O \citep{maeder92} and 0.2-0.8\Msol~of Fe (see discussion in \citealt{silich01}). Assuming a number density of 0.01-0.2 \pcc, spherical geometry, a radius of 50 pc, and 10\% He abundance, the total mass enclosed by the bubble should be between 100 and 3000\Msol. As the inferred number density from the observed spectrum (see Equation~\ref{eqn:dens}) is 0.2\pcc, we assume an enclosed mass of 3000\Msol~in what follows. We use Equations 9 and 10 from \citet{silich01} to calculate the resulting metallicity:

\begin{equation}
\label{eqn:silichO}
Z_{\rm O} = \frac{M_{\rm ej,O}/Z_{\rm sol,O}+Z_{\rm ISM}M_{\rm ev}}{M_{\rm ev}+M_{\rm ej}}
\end{equation}

\begin{equation}
\label{eqn:silichFe}
Z_{\rm Fe} = \frac{M_{\rm ej,Fe}/Z_{\rm sol,Fe}+Z_{\rm ISM}M_{\rm ev}}{M_{\rm ev}+M_{\rm ej}}
\end{equation}

$M_{\rm ej}$ is the mass ejected in SNe, $M_{\rm ev}$ is the mass evaporated from the shell walls, and $Z_{\rm sol,O}$ and $Z_{\rm sol,Fe}$ are the solar abundances by mass from \citet{grevesse96}, $Z_{\rm sol,O} = 0.0083$ and $Z_{\rm sol,Fe} = 0.00126$. \citet{weaver77} predict that the evaporated mass will be the dominant source of mass in the bubble interior. We assume that, apart from the mass ejected by the four SNe, the remainder of the mass was contributed by evaporation. Assuming stellar remnant masses of $\sim2$\Msol~\citep{maeder92} and the initial masses from \citet{oey95}, the total mass returned to the ISM by the four most massive stars is 257\Msol, resulting in a $M_{\rm ev}$ of $\sim$~2700\Msol. Then, from equations \ref{eqn:silichO} and \ref{eqn:silichFe}, the predicted metallicities are $Z_{\rm O}=3\Zsol$ and $Z_{\rm Fe}=0.4-0.6\Zsol$. If the bubble contains less evaporated mass, these metallicities will be higher. 

As noted by \citet{silich01}, the X-ray emissivity is proportional to the metallicity. Assuming the $\alpha$ elements dominate the soft X-ray emission, the factor of 10 increase from $Z_{\rm LMC}$ of 0.4\Zsol~to $Z_{\rm O}$ of 3\Zsol~would be sufficient to explain the factor of 10 increase in DEM L152's luminosity (see Table~\ref{tab:models}.)  This predicted metallicity does not match the observed $\alpha$ abundance in DEM L152, however (see Table~\ref{tab:fit152}), which is subsolar in all cases with reasonable error bars. The $\alpha$/Fe ratio is a more reliable diagnostic of the metallicity, and it is possible that the $\alpha$ abundance from the spectral fits is incorrect. Given the additional uncertainties in the bubble's density, geometry, and resulting mass and in the number of SNe that have occurred, this discrepancy between the predicted and observed metallicities is not unexpected. The predicted $\alpha$/Fe ratio from the 4 SNe is 5-8, while Table~\ref{tab:fit152} shows typical $\alpha$/Fe ratios of around 3. When errors are taken into account, the observed $\alpha$/Fe ratio could be consistent with the predicted value, and therefore it is possible that the higher $\alpha$ abundance is also present. 

If we use the observed $\alpha$ abundances, which have an average value of $\sim~0.6$\Zsol~and are generally consistent with 1\Zsol, the enhancement over the Weaver model only accounts for 14-20\% of the observed emission. At most, the observed $\alpha$ abundance can account for the diffuse emission in the Bubble and West regions and the West Blowout, but not for the emission from the Bubble Limb, West Limb, or South Bar. If, on the other hand, we accept that this observed $\alpha$ abundance is inaccurate and use the predicted $Z_{\rm O}$, the enhanced metallicity explains the X-ray luminosity but not the limb-brightened features or enhanced expansion velocities. 

A similar analysis for DEM L50 yields a predicted $Z_{\rm O}$ of 5\Zsol~and predicted $\alpha$/Fe of 7-13, where we have assumed $M_{\rm ej,O}=36$\Msol, $M_{\rm ej,Fe}=0.1-0.4$\Msol, a bubble radius of 50 pc, and $n_{\rm H}=0.06$\pcc, derived from the non-equilibrium fit normalization. The predicted metallicities are much higher than the approximately solar values obtained from the non-equilibrium fit, and as with DEM L152, the observed metallicities cannot account for the majority of the X-ray enhancement. 

\subsection{Mass-loading by clouds}
\label{sec:discussion:massloading}

Another possible explanation for the high X-ray luminosities of DEM L50 and DEM L152 is mass loading from the evaporation or ablation of overrun clouds. While the distribution of mass from evaporation at the shell walls should be uniform over relatively large scales, evaporating clouds should appear as small clumps of emission in our data. DEM L50 is located in a void and does not exhibit any noticeable clumpiness in its emission. The two knots we observed in DEM L152 and its more crowded HI environment, on the other hand, show that mass-loading by clumps could be occurring in that bubble. \citet{silich96} investigate clouds with $n_e\sim10$\pcc and radii of 3-5 pc, while \citet{orlando05} find their modeled shocked clouds have number densities of 4\pcc, radii of 1 pc, and are surrounded by a corona with density 0.4\pcc. To encompass this range of possible cloud parameters, we calculated the emission from clouds with each of these three densities ($n_e$=0.4, 4, and 10 \pcc), a temperature of $10^{6}$ K, a metallicity of 0.4\Zsol, and radii of 0.5 pc, 1 pc, and 5 pc, all of which should be resolved at Chandra's $\sim1$" resolution. The X-ray emissivity in the 0.2-4.5 keV band is approximately $\Lambda_x=9\times10^{-24}\times Z$ erg \ps \pcc \citep{silich96}; this emissivity will result in more emission than we expect to see in the 0.3-2.0 keV band, but it should allow us to roughly estimate the type of clouds that could be detectable. Assuming a median photon energy of 0.55 keV \citep{orlando10}, we can calculate the predicted count rate per arcsec$^2$ (Table~\ref{tab:knotcts}). A comparison with the X-ray contours in Figures~\ref{fig:img_deml50} and~\ref{fig:img_deml152} shows that the knots with radii of 0.5 pc are potentially detectable if they have a density of $\sim10$\pcc, while no knots with a density of $\sim0.4$\pcc are detectable. Interestingly, the detected ``knots" in DEM L152 have a count rate of approximately $6.4 \times 10^{-6}$ s$^{-1}$ arcsec$^{-2}$, which is within the range of values in Table~\ref{tab:knotcts} and a density of $\sim1$\pcc, found using Equation~\ref{eqn:dens}, assuming a radius of 2.5 pc. These values suggest that these knots may indeed be genuine clouds swept up by the bubble.

Another possibility is that some clumps appeared as point sources in the images (Fig.~\ref{fig:point_sources}). We would expect  predominantly soft emission from overrun clumps, while other point sources, such as background AGN or X-ray binaries, should appear in the hard band as well. Only two point sources, one in DEM L50 and one in DEM L152, appear in the soft band alone. The point source in DEM L50 is quite faint, containing only $\sim15$ counts, and its identity is unclear. The soft point source in DEM L152 is located outside the main bubble. This source is likely associated with the highly ionized N44 C region rather than an independent cloud swept up by DEM L152. In any case, the lack of multiple soft point sources suggests that dense, overrun globules are not present to any significant degree in either bubble. 

\begin{deluxetable}{llll}
\tabletypesize{\scriptsize}%
\tablecolumns{4}
\tablewidth{0pc}
\tablecaption{Predicted Cloud Count Rates
        \label{tab:knotcts}}
\tablehead{
\colhead{$n_e$(\pcc)} & \colhead{$R_{\rm cl}$=0.5 pc} & \colhead{$R_{\rm cl}$=1.0 pc} & \colhead{$R_{\rm cl}$=5.0 pc}
  \\
\colhead{(1)} & \colhead{(2)} & \colhead{(3)} & \colhead{(4)} 
  }
\startdata
0.4 & $2.5 \times 10^{-9}$ & $5.0 \times 10^{-9}$ & $2.5 \times 10^{-8}$\\
4.0 & $2.5 \times 10^{-7}$ & $5.0 \times 10^{-7}$ & $2.5 \times 10^{-6}$\\
10.0 & $1.5 \times 10^{-6}$ & $3.1 \times 10^{-6}$ & $1.5 \times 10^{-5}$\\
\enddata
\tablecomments{
  Count rates are given for the 0.2-4.5 keV energy band in photons s$^{-1}$ arcsec$^{-2}$. 
}
\end{deluxetable}

Although DEM L50 and DEM L152 may have overrun additional clouds which are below our detection limit or already destroyed, their morphology and spectral properties suggest that mass-loading by clouds is not dominating the bubbles' evolution. Modeled bubbles with high levels of mass loading tend to show a smoother distribution of density and temperature across the bubble, as the clouds add cooler gas to the system \citep[see \eg][]{arthur96,pittard07}. The increased central density and reduced central temperature may result in a flat surface brightness profile \citep{arthur96} or even a centrally-peaked profile \citep{silich96}. A limb-brightened morphology, on the other hand, appears correlated with supernova activity. No pre-supernova superbubbles yet studied show limb-brightening, while this morphology is common in post-supernova superbubbles \citep{chu_rmxac}. DEM L50 and DEM L152's limb-brightened morphologies likely reflect their supernova history rather than a history of high mass-loading. 

Mass-loading should reduce the central temperature of a bubble \citep[\eg][]{silich96}, whereas DEM L152's central temperature is quite high ($\sim~0.37$ keV), higher than predicted by the Weaver model. DEM L50's lower central temperature of 0.15 keV in the equilibrium fit could be consistent with mass-loading, although uncertainties in the number of SNe or energy input in DEM L50 could also account for the slightly lower temperature. The temperature derived from the non-equilibrium fit to DEM L50 is completely consistent with the Weaver model predictions, however, requiring no additional mass-loading. Mass-loading by overrun clouds or evaporation from the shell walls is likely occurring to some extent in both bubbles, but it does not fully explain the bubbles' X-ray emission or large-scale properties. 

Dense cloudlets have been detected in other superbubbles, but their presence does not necessarily imply a high X-ray luminosity. For instance, although clouds are detected in the superbubble DEM L192 and the bubble appears to be expanding into an inhomogeneous ISM, its X-ray luminosity in the 0.3-3.0 keV band is only moderate, $\sim4\times10^{35}$ \ergps \citep{cooper04}. \citet{cooper04} note a correlation between the \halpha~and X-ray surface brightnesses in this bubble, however, and suggest that mass-loading is occurring through evaporation at the shell walls. This analysis is consistent with our findings in this section and in \S~\ref{sec:models:hd}; thermal conduction and the subsequent evaporation of shell material is important in DEM L50 and DEM L152, while the contribution of swept-up clouds to the X-ray emission is negligible. 

Thus, we see evidence that both metallicity enhancement and mass-loading by clouds has taken place in these superbubbles. The metallicity enrichment is too low to account for the observed luminosities, however. Likewise, few candidate clouds are observed in the superbubbles and the X-ray emission is inconsistent with a high mass-loading scenario. Metallicity and mass-loading by clouds do not completely explain the observed X-ray luminosities.

\section{Conclusions}
\label{sec:conclusion}

We present high spatial resolution {\it Chandra} ACIS-S images and spectra of the diffuse emission in two X-ray-bright LMC superbubbles, DEM L50 and DEM L152. These objects have X-ray luminosities on the order of $10^{36}$ \ergps, an order of magnitude higher than predicted by the standard \citet{weaver77} wind-blown bubble model. The level of detail in the {\it Chandra} observations allows us to evaluate the relative importance of the various mechanisms that contribute to the X-ray luminosity by constraining the spatial origin of the emission and revealing X-ray spectral variations across the superbubbles.

With these high-quality images, we isolated the diffuse X-ray emission by excluding point sources and examined the spectra of separate regions within each superbubble. X-ray images of both bubbles show limb-brightened morphologies. Most of the limb-brightening in DEM L50 is the result of a higher central absorption column density, while the limb-brightening at the bubble's southern end may be real. Likewise, the intervening absorption does not account for the limb-brightening in Shell 1 of DEM L152. Hardness ratios indicate that these same regions, the Bright Limb subregion of DEM L50 and Shell 1 of DEM L152, are spectrally harder than the other regions in the superbubbles. 

Spectral fits to the superbubbles account for the variations in absorption column density from region to region and reveal the temperature differences and luminosity contribution of each region. The spectra of both objects show evidence of SNR impacts on the shell walls. In particular, DEM L50's Bright Limb has a harder spectrum, higher temperature, and higher surface brightness than the rest of the bubble. These spectral properties support the idea that a SNR recently heated the Limb region. The limb-brightened Shell 1 region in DEM L152 also has a higher temperature and surface brightness compared to other regions of the superbubble. A high temperature component in the limb-brightened West region and the adjacent South Bar blowout may be another indicator of recent heating due to SNe.

Our spectral fits also revealed unexpected abundance ratios in DEM L50, which suggest the superbubble is not in collisional ionization equilibrium. DEM L50's low $\alpha$/Fe ratio is the opposite of the expected $\alpha$-enrichment from core-collapse SNe. While we cannot rule out the possibility of a Type Ia SN in the area, the consistency of the ratio across the bubble suggests that either the bubble happens to lie in an unusual ISM environment or, more likely, DEM L50 is not in ionization equilibrium. Assuming a non-equilibrium situation, the derived ionization timescales are on the order of 10$^5$ years, which is consistent with the other indications of a recent supernova in the bubble. 

We compared our observations with both the standard wind-blown bubble model \citep{weaver77} and a one-dimensional hydrodynamic model based on \citet{oey04b}. Both models include the time-dependent energy input of the stellar population. As shown by the Weaver model predicts a total X-ray luminosity an order of magnitude below the observed value. While the 1-D hydrodynamic simulation reproduces the size and expansion velocity of DEM L50 by assuming a higher ambient pressure and including the effects of SNR impacts, the neglect of thermal conduction produces X-ray luminosities far below our observed luminosities. This discrepancy indicates that thermal conduction is important in these bubbles, as evaporation of material from the shell walls enhances the density of the interior. The SNR impact does significantly brighten the modeled bubble, particularly at the limb, which supports the off-center SNR explanation for the existence of X-ray-bright superbubbles \citep{chu90}. 

Our observations do not reveal signs of significant metal enrichment, an alternative explanation for the high observed luminosities \citep{silich01}. A higher proportion of metals could enhance the X-ray emissivity, but neither superbubble exhibits a sufficiently high metallicity to account for the anomalous X-ray luminosity. Abundances determined from X-ray observations are uncertain, however, and the bubbles may be more enriched than the spectral fits suggest. While a high metallicity could raise the luminosity, it would still not explain the observed morphologies or large expansion velocities of the objects. 

We consider the hypothesis that mass-loading through the ablation or evaporation of swept-up clouds could increase the density of the superbubbles and raise the X-ray luminosity \citep[\eg][]{mckee84,pittard01}. With {\it Chandra}'s $\sim$~1" resolution, we expect to detect clouds a few parsecs in diameter. Few candidate clouds are observed, and they contribute negligible emission to the superbubbles. Although some mass-loading is likely present and the superbubbles may have overrun more clouds in the past, the effect of clouds on the X-ray emission is insignificant compared to the luminosity enhancement from SNR impacts.

An examination of the energy budget for both superbubbles further clarifies the role of off-center SNRs. The relatively small Bright Limb subregion of DEM L50 generates roughly a quarter of the total luminosity, while the limb as a whole accounts for half of the total X-ray luminosity. In DEM L152, the Shell 1 Limb and West Limb regions alone provide about half of the total X-ray luminosity. All these regions exhibit signs of supernova activity. On the other hand, the contribution of metallicity enrichment and mass-loading from clouds to the X-ray luminosity is much less. Metallicity enrichment accounts for at most 20\% of the excess emission, while dense clouds account for less than 2\% of the total emission. SNR impacts and evaporation of material from the shell walls dominate the observed X-ray luminosities.

Our {\it Chandra} images and spectra of DEM L50 and DEM L152 are the first observations of X-ray-bright superbubbles at a sufficiently high spatial and spectral resolution to distinguish between various emission mechanisms. While mass-loading and metallicity enrichment due to SNe must be occurring to some degree, our observations confirm that thermal evaporation from the shell walls and off-center SNR impacts are the primary sources of the enhanced X-ray luminosity. SNR impacts increase the shell expansion velocity, lead to higher radiative losses, and may contribute to blowouts and the subsequent mixing of ISM and hot, enriched gas. Although the X-ray luminosities are not high enough to explain the superbubble growth-rate discrepancy, X-ray radiative losses will decrease the energy available for transport to the ISM. Unlike the cloud mass-loading scenario, SNR impacts and thermal conduction should affect most superbubbles at some point in their evolution, irrespective of their environment. Off-center SNR impacts in superbubbles play a critical role in the evolution of the ISM by enhancing the mixing of hot, enriched gas with ambient material, while at the same time leading to X-ray cooling that decreases the available energy.

\acknowledgments{We thank Robert Gruendl and Lister Staveley-Smith for their assistance in extracting the \hi~column density maps and Colin Slater and Mathieu Compi{\`e}gne for helpful discussions about dust properties. We also thank an anonymous referee for useful suggestions and Robert Gruendl for help with the median energy analysis. We acknowledge support from NASA grant GO2-3193C.}

\end{document}